\documentclass[conference,review]{IEEEtran}
\IEEEoverridecommandlockouts

\usepackage{listings}
\usepackage{adjustbox}
\usepackage{float}
\usepackage{graphics}
\usepackage{xcolor}
\usepackage{varwidth}
\usepackage{multicol}
\usepackage{stfloats}
\usepackage{color, colortbl} 
\usepackage{cite}
\usepackage{amsmath,amssymb,amsfonts}
\usepackage{algorithmic}
\usepackage[linesnumbered,lined,boxed,ruled,vlined]{algorithm2e}

\usepackage{svg}
\usepackage{hyperref}

\usepackage{array, boldline, rotating} 
\usepackage{color, colortbl} 
\definecolor{tabcol}{rgb}{0.7,0.8,1}
\usepackage{booktabs}   

\usepackage{lipsum}

\newcommand{\DDT}{DDF}

\definecolor{commentgreen}{RGB}{2,112,10}

\SetCommentSty{mycommfont}

\usepackage{graphicx}
\usepackage{textcomp}
\usepackage{xcolor}
\def\BibTeX{{\rm B\kern-.05em{\sc i\kern-.025em b}\kern-.08em
    T\kern-.1667em\lower.7ex\hbox{E}\kern-.125emX}}

\begin{document}

\title{Vectorizing Sparse Matrix Codes with Dependency Driven Trace Analysis
}

\author{\IEEEauthorblockN{Zachary Cetinic}
\IEEEauthorblockA{
\textit{University of Toronto}\\
Toronto, Canada \\
zachary.cetinic@mail.utoronto.ca}
\and
\IEEEauthorblockN{Kazem Cheshmi}
\IEEEauthorblockA{
\textit{University of Toronto}\\
Toronto, Canada \\
kazem@cs.toronto.edu}
\and
\IEEEauthorblockN{Maryam Mehri Dehnavi}
\IEEEauthorblockA{
\textit{University of Toronto}\\
Toronto, Canada \\
mmehride@cs.toronto.edu}
}
\maketitle

\definecolor{commentgreen}{RGB}{2,112,10}
\definecolor{grayLine}{RGB}{100,100,100}
\definecolor{annotation}{rgb}{0.45, 0.31, 0.59}
\definecolor{darklavender}{rgb}{0.45, 0.31, 0.59}
\definecolor{asparagus}{rgb}{0.12, 0.3, 0.17}
\definecolor{brightmaroon}{rgb}{0.76, 0.13, 0.28}
\definecolor{bondiblue}{rgb}{0.0, 0.58, 0.71}
\lstset{ %
language=C++,                
basicstyle=\ttfamily\footnotesize,       
columns=fullflexible,
numbers=left,                   
numberstyle=\footnotesize,      
stepnumber=1,                   
numbersep=5pt,                  
showspaces=false,               
showstringspaces=false,         
showtabs=false,                 
frame=none,           
tabsize=2,          
captionpos=b,           
breaklines=true,        
breakatwhitespace=false,    
escapeinside={\%*}{*)},          
keywordstyle=\color{blue},       
commentstyle=\color{commentgreen}\ttfamily,
otherkeywords={atomic,...},           
numberstyle=\tiny\color{black}, 
rulecolor=\color{black},
escapechar=|
}


\begin{abstract}
Sparse computations frequently appear in scientific simulations and the performance of these simulations rely heavily on the optimization of the sparse codes. The compact data structures and irregular computation patterns in sparse matrix computations introduce challenges to vectorizing these codes. Available approaches primarily vectorize regular regions of computations in the sparse code. They also reorganize data and computations, at a cost, to increase the number of regular regions.  In this work, we propose a novel polyhedral model, called the partially strided codelets (PSC), that enables the vectorization of computation regions with irregular data access patterns. PSCs also improve data locality in sparse computation. Our \DDT{}  inspector-executor framework efficiently mines the memory accesses in the sparse computation, using an access function differentiation approach, to find PSC codelets. It generates vectorized code for the sparse matrix multiplication kernel (SpMV), a kernel with parallel outer loops, and for kernels with carried dependence, specifically the sparse triangular solver (SpTRSV).
We demonstrate the performance of the \DDT{}-generated code on a set of 60 large and small matrices (0.05-130M nonzeros). \DDT{} outperforms the highly specialized library MKL  with an average speedup of 1.93 and 4.5$\times$ for SpMV and SpTRSV, respectively. For the same matrices, \DDT{} outperforms the state-of-the-art inspector-executor framework Sympiler \cite{Cheshmi2017} for the SpTRSV kernel by up to 11$\times$ and the work by Augustine et. al \cite{Augustine2019} for the SpMV kernel by up to 12$\times$. 

%
%
\end{abstract}


\section{Introduction}

Sparse matrix computations are important kernels used in a large class of scientific and machine learning applications. The performance of  sparse kernels is noticeably improved if the code is \textit{vectorized} to exploit single instruction multiple data (SIMD) capabilities of the underlying architecture. 
%
Vectorization potentially increases opportunities to optimize for locality, further increasing the performance of the sparse code. SIMD instructions can efficiently vectorize   groups of operations that access consecutive data, i.e. have a regular access pattern.  A computation is regular when its operations access memory addresses with a constant distance from each other, i.e. \textit{strided access}, and otherwise irregular.   However, because a sparse matrix  is typically stored in a compact representation \cite{bulucc2009parallel}, the sparse matrix code accesses are often irregular, making vectorization challenging.  
Not vectorizing irregular segments in the sparse matrix calculation  can potentially reduce locality and the performance of the sparse matrix kernel.

The most common approach to improve vectorization of irregular regions in sparse matrix codes is to reorganize data or computations so the region becomes regular and amenable to vectorization. These work typically optimize sparse kernels with no loop-carried dependencies, for example  
\cite{intel-alt,Kreutzer2013,Liu2013,Liu2015,Tang2015,Chen2016,Xie2018} improve the performance of sparse matrix-vector multiplication (SpMV) with reorganization. CSR5~\cite{liu2015csr5} and CVR~\cite{xie2018cvr} reorganize data to create contiguous accesses and reorganize computations to increase the number of independent operations for vectorization. However, their computation reorganization typically increases the number of floating-point operations in the kernel, reducing the benefits of vectorization. 
Sparse storage formats such as ELL and DIA~\cite{saad2003iterative} are used to pad the sparse matrix with additional nonzeros. Padding increases regular regions in the matrix and hence increases the number of consecutive vector loads. In sparse matrices with very irregular patterns, the amount of padding will increase, leading to more floating-point operations.   
 Reorganization approaches  are often library-based and thus are specialized for a specific set of matrix patterns or the access patterns appearing in a specific matrix. They also need to be manually ported to new architectures.
For example, work in \cite{Yesil2020} primarily optimizes matrices from power-law graph computations, CVR is for sparse graphs, and ELL is most efficient for block-structured matrices such as factors of a direct solver~\cite{vazquez2010improving}. 

Compilers, such as those based on the polyhedral model,  generate vectorizable and portable code for affine codes such as dense matrix computations \cite{Boulet1998,Chen2012,Kershaw1978,Puschel2005,Spampinato2014,Tiwari2009}. However, compilers are limited in vectorizing sparse codes because of the existing   indirection in the memory access patterns of sparse computations, caused by data compaction. This indirection introduces additional challenges to compilers when optimizing sparse kernels with loop carried dependencies, e.g. the sparse triangular solver. Recently the polyhedral method was extended to optimize sparse matrix computations via inspector-executor approaches \cite{strout2012set,Venkat2015,strout2018sparse}. The  inspectors execute at runtime to resolve indirection and find data  dependencies  between  operations and this information is used  to transform the original source code into optimized  code called the executor. 


\begin{figure*}[!ht]
    \includegraphics[width=\textwidth]{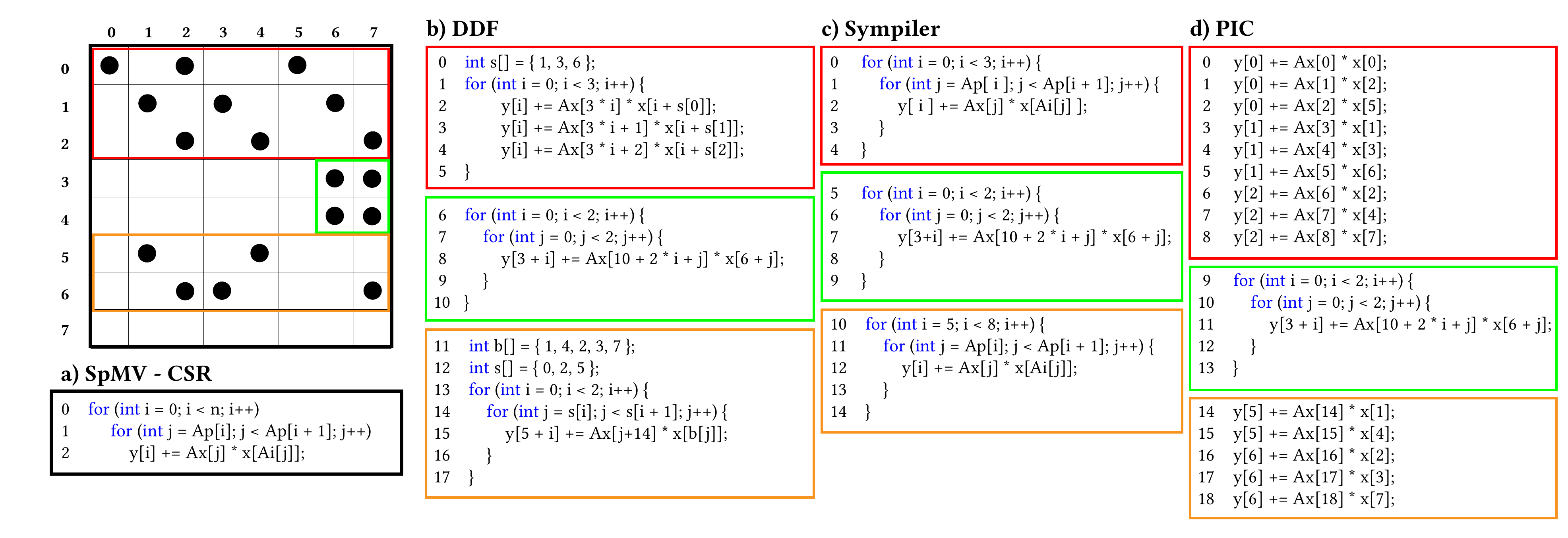}
    \caption{ Various generated codes for the given sparsity pattern (shown in the matrix) used to compute the Sparse Matrix-Vector multiplication (SpMV) kernel. \textbf{a)} Shows the code that computes SpMV using the compressed sparse row (CSR) storage format. \textbf{b)} Shows the \DDT{} generated code.  \textbf{c)} Shows the code generated from the Sympiler framework \cite{Cheshmi2017}. \textbf{d)} Shows the code generated using the work by Augustine \textit{et al.}\cite{Augustine2019}}
    \label{overview}
\end{figure*}

Amongst inspector-executor frameworks, Augustine et al.~\cite{Augustine2019}, which we refer to as the PIC framework, and Sympiler ~\cite{Cheshmi2017} inspect the sparsity pattern of input matrices to generate vectorized code for sparse matrix computations.   The work in \cite{Augustine2019} proposes an inspector to mine the memory trace of the SpMV kernel to find regular regions in the computation for vectorization. The remaining regions are unrolled. Unrolling irregular regions limits the scalability of this work because of increasing code size for large matrices and for matrices with high irregularity; their framework can primarily optimize small sparse matrices (with less than 0.5 million nonzeros). Also, they do not support sparse kernels with loop-carried dependencies. Sympiler uses an inspector to resolve data dependencies in sparse kernels such as SPTRV.  Its inspector finds iterations that operate on rows with similar patterns, i.e. regular regions, and applies code transformations that primarily optimize for tiling and vectorization.  Sympiler adds additional nonzeros to some of the rows with non-consecutive memory accesses and thus pads the input matrix to form \textit{row-blocks}. Row-blocks are regular regions that can be easily vectorized.  To limit the overheads from padding, Sympiler only creates row-blocks when profitable, thus limiting the size of row blocks it creates. It is thus not able to efficiently optimize sparse matrices with few rows of similar sparsity patterns.

The standard approach of prior work in vectorizing sparse kernels is to 
either find regions with regular computations or create more regular regions in the computations with for example data reorganization. The focus of this work is to expose vectorization opportunities in computation regions with irregular computation patterns. We present a novel polyhedral model which we call the \textit{partially strided codelet} that increases opportunities for vectorization of sparse matrix codes and improves data locality. PSCs represent computation regions that have non-consecutive, i.e. unstrided, memory accesses but can still benefit from vectorization because a subset of their accesses is strided.  
We also present a novel inspector-executor framework called \DDT{} (Differentiating Data access Functions) that efficiently inspects the memory access patterns in sparse matrix codes to mine for PSCs using an \textit{access function differentiation} approach and to generate optimized vectorized code for the matrix computation. \DDT{}'s inspector supports sparse kernels with and without loop-carried dependencies as it is able to look for data dependencies in the computation region. 



The contributions of the work are:
\begin{itemize}

\item A novel polyhedral model, called the partially strided codelet, that enables the vectorization of computation regions with non-consecutive memory accesses in sparse codes while improving temporal and spatial locality.

\item A memory access differentiation approach, that uses the \textit{first order partial difference (FOPD)} of access functions to distinguish strided access from non-strided accesses in the sparse matrix computation. FOPD's are used to mine for PSCs in spare matrix codes. 

\item The \DDT{} inspector-executor framework that mines the data accesses of sparse codes with or without dependencies to find PSCs. The inspector executes in parallel with low overhead and generates a compact code that can execute in parallel. 

\item We demonstrate the performance of \DDT{} for SpTRSV and SpMV kernels. 
%
The speedup of the \DDT{}-specialized code for SpTRSV is on average 1.79$\times$ and 4.5$\times$ faster than that of Sympiler and MKL respectively. For SpMV, \DDT{}-generated code is on average 1.42$\times$ and 1.93$\times$ faster than that of CSR5 and MKL respectively. For small problems that the work from \cite{Augustine2019} supports, \DDT{} is 10.57$\times$ faster than their approach. 
\end{itemize}


 
 \section{Motivation}
In this section, we use the matrix in Figure~\ref{overview} to explain the approach used in inspector-executor frameworks, i.e. Sympiler and the work in ~\cite{Augustine2019} (which we call PIC), to optimize different computation regions in the SpMV kernel and compare to the approach proposed in this work. The example uses the compressed sparse row (CSR) format to store matrix $A$ and then computes $y= A\times x$ where $x$ is a vector.  The region with the green border is a regular computation region in the SpMV computation and the red and orange colors show regions with non-strided accesses. 


All tools can efficiently vectorize the green region 
 since all of its accesses are consecutive and thus strided. BLAS routines are typically used to optimize such regions. They enable vectorization, and in part, also improve locality by reusing the entries in the input vector (e.g. \texttt{x[6]}, \texttt{x[7]}) and the output vector (e.g. \texttt{y[3]}, \texttt{y[4]}). 

For the non-strided segments, Sympiler uses a padding strategy, when profitable, to create regular regions that can be easily vectorized. However, Sympiler will not pad the red and yellow regions, since they are too irregular, and padding will introduce a large number of additional operations.  Instead, Sympiler will fall back to the original non-affine code and miss any additional vectorization opportunities. 
PIC will generate fully unrolled code for the non-strided regions, as shown in Figure~\ref{overview}d. While this reduces the number of instructions, it increases code size and as a result, increases the number of instruction cache misses, especially for large matrices, limiting PIC's scalability. 

For irregular regions, \DDT{} uses partially strided codelets to improve locality and to provide a better vectorization efficiency. 
The SpMV code for the red region
has non-strided accesses in \texttt{x}, while its accesses to \texttt{Ax} and \texttt{y} are strided.  \DDT{} generates the PSC codelet bordered with  red in Figure~\ref{overview}b. The code stores the non-strided addresses in \texttt{s} to be reused across all three iterations of $i$ instead of loading indices for every operation. This also enables the reuse of index \texttt{Ap[i]} in \texttt{Ax} so it does not have to be loaded per each outermost iteration. 
Similarly, for the yellow region, accesses to both \texttt{Ax} and \texttt{x} are non-strided, but vectorizing computations of the innermost loop improves spatial locality due to contiguous accesses in \texttt{Ax}.

\section{Partially Strided Codelet}
This section introduces partially strided codelets and discusses their classification as well as efficiency in optimizing computation regions with irregular access patterns. We also present a novel cost model and a differentiation-based PSC detection strategy, both of which are used in the \DDT{} framework to efficiently find PSCs in sparse matrix computations. 



\begin{figure}[t]
    \centering
    \includegraphics[width=\linewidth]{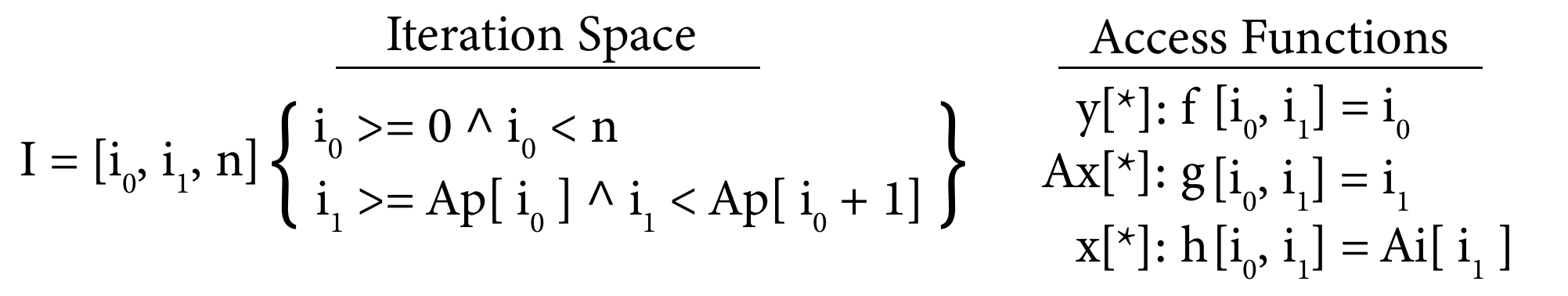}
    \caption{Polyhedral representation of the SpMV kernel shown in Figure~\ref{overview}a}
    \label{fig:polyhedral_representation}
\end{figure}

\subsection{Definitions}
\label{sec:defs}


\textit{Polyhedral model and data access functions.} 
A loop nest that contains a set of statements is represented with a polyhedral model through an integer polyhedron sets $\mathcal{I}$ and relations $f$.
A statement is made  of a data space described with $\mathcal{D}$ which is a disjoint set containing $\mathcal{D}_0,\ldots,\mathcal{D}_n$. 
An integer polyhedral set $\mathcal{I}=[i_0,...,i_n]$ is a collection of inequalities that create bounds for each dimension inside $i
\in\mathcal{I}$. 
For each $\mathcal{D}_d\in \mathcal{D}$  a \textit{data access function} $f$ is used to describe how the data space $\mathcal{D}_d$ is accessed by the iteration space of $\mathcal{I}$. In other words, a data access function maps an iteration space to a data space, i.e. $f_{\mathcal{I}\rightarrow\mathcal{D}}$ . 
The SpMV code in Figure~\ref{overview}a has one statement. That statement has three data spaces \texttt{y}, \texttt{Ax}, and \texttt{x} as well as three data access functions. Figure~\ref{fig:polyhedral_representation} shows the polyhedron sets for $\mathcal{I}=[i_0, i_1]$ and also the access functions corresponding to each data space.

\begin{figure}[t]
    \centering
    \includegraphics[width=\linewidth]{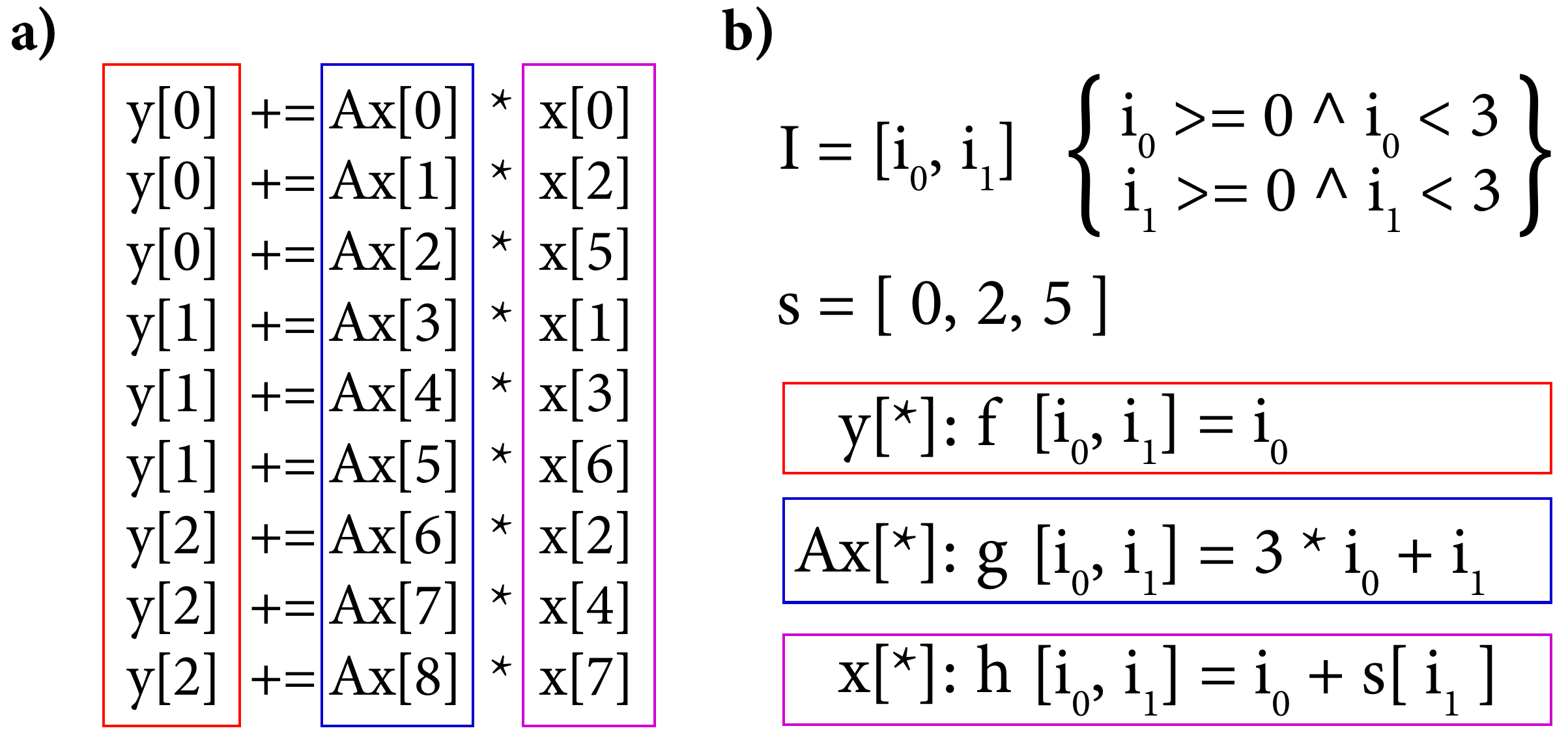}
    \caption{\textbf{a)} Shows the set of SpMV operations from the red outlined region of the matrix in Figure \ref{overview}. \textbf{b)} Shows the iteration space and access functions of a codelet that maps to the data-space seen in section $a)$ of the figure.}
    \label{fig:dataspace_example}
\end{figure}

\textit{Codelet.} A polyhedral model that has a convex integer polyhedron with no flow dependencies, i.e. read after write dependency between access functions, is a codelet.
A codelet only has one statement and the operation in that statement is a SIMD-supported operation. 
A set of SpMV operations is shown in Figure~\ref{fig:dataspace_example}a and its polyhedral representation including iteration space and data access functions are shown in Figure~\ref{fig:dataspace_example}b. All operations in this model are independent and have only one statement (y[*] += Ax[*] * x[*]), with a multiply-add operation which is supported by Intel and AMD SIMD units. These properties satisfy the criteria for being a codelet.  


\textit{Strided data access function.} A function that can be expressed with a linear combination of induction variables in $\mathcal{I}$ is a strided access function. The data access function $f$ and $g$ in Figure~\ref{fig:dataspace_example}b are both strided and can be expressed as a linear combination of $[i_0,i_1]$.


\subsection{Partially Strided Codelet Classification}
\label{sec:pscclas}

An efficient way to vectorize a codelet is to find operations with strided accesses across different iterations. This enables the vectorization of more operations and potentially increases data reuse between different iterations.
However, current approaches are primarily limited to  vectorizing codelets that  all of their access functions are strided. An efficient BLAS~\cite{blackford2002updated} implementation is used for this purpose and thus, we call these codelets types  BLAS codelets. In this work we define a novel set of codelets called \textit{Partially Strided Codelets} that have  
 at most $n-1$ strided access functions and at least one non-strided access function. These codelets can benefit from  vectorization because they have one or more strided access functions. 

 PSCs are classified into different types based on the number of access functions that are strided. For example in SpMV and SpTRSV that have codelets with three access functions, $n=3$, two types of PSCs can be defined. The PSC type I codelet is used when two of access functions are strided, and the PSC type II codelet is used when only one access function is strided.  Figure~\ref{overview}b shows two types of PSCs, outlined in yellow and red, and also the BLAS codelet, outlined in green.

\subsection{PSC Efficiency and Cost Model}
\label{subs:cost}


\begin{figure}[!t]
    \centering
    \includegraphics[width=\linewidth]{{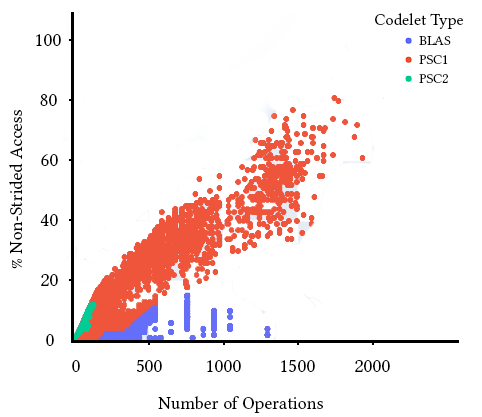}}
    \caption{Shows best performing codelet type on a wide variety (150K total points) of computational regions found in SpMV/SpTRSV.}
    \label{fig:codelet_observations}
\end{figure}

Depending on the number of operations and strided accesses in a computation region, either PSC or BLAS codelets can be profitable.  
To illustrate this, we extracted over 150,000 computation regions, with a different number of operations and strided accesses, from the SpMV and SpTRSV kernels on 60 matrices in the Suitesparse repository \cite{davis2011university}.  Each region is optimized using one or several codelets of the same type and the best performing codelet type for that region is inserted as a point in Figure~\ref{fig:codelet_observations}. The red, green, and blue colors are used for  PSC I, PSC II, or BLAS codelets respectively. The figure shows that some regions perform best with PSCs and some with BLAS codelets, e.g. PSC I provides the fastest execution time over all codelets for regions with a large percentage of non-strided accesses. 

The performance of codelets differs based on the number of memory accesses that occur from running the codelets.
Per Section~\ref{sec:defs}, an access function is described as $f(\mathcal{I})=x + s[i_0,  \ldots, i_n]$ where $x$ is an integer and $s$ is an $n$-dimensional array.
If $f$ is a strided access function, an integer $s[i]$ for each $i\in\mathcal{I}$  needs to be loaded from memory before the operation in the codelets executes. Each integer, $s[i]$, is the $i^{th}$ coefficient in the linear combination of $f$.  
However, for a non-strided access function, the number of memory accesses in the codelet is equal to the size of $s[i]$ for $i\in\mathcal{I}$, which is larger than one. Thus, compared to BLAS codelets, PSCs access more memory addresses. But this does not imply that PSC codelets are inefficient because 
the number of codelets used to cover a computational region also affects the total memory accesses. 
%
%
For example, the computation region in  Figure \ref{fig:dataspace_example}b is a PSC I with one non-strided access function $h(i_0, i_1) = i_0 + s[i_1]$ where $s=\{0,2,5\}$ and thus requires loading 4 memory addresses. If BLAS is selected for the same region, a minimum of 3 codelets will need to be used leading to 9 loads. 

\begin{figure}[!t]
    \centering
    \includegraphics[width=\linewidth]{{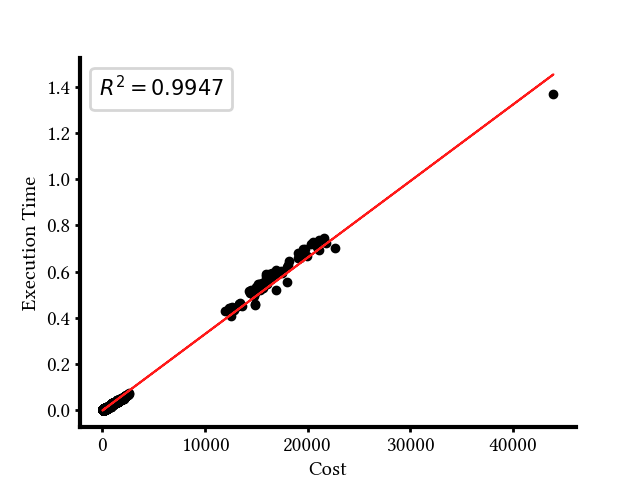}}
    \caption{Correlation between predicted cost of codelets and average execution time of each codelet. Averages were obtained using 1,000,000 executions per codelet.}
    \label{fig:cost_correlation}
\end{figure}


To efficiently determine the types of codelets for a computation region, we first define a cost model to compute the cost of a single codelet.  
%
The cost model uses the number of memory accesses of each memory access function, i.e. $|s_d|+1$ and the number of operations in a codelet($|p|$) to estimate the execution time of the codelet. 
The cost of a codelet $l$ is $c_l$  defined as: 
\begin{equation}
    c_l = |p| + m + \sum_{d=0}^{m}|s_d|
    \label{eq:cost}
\end{equation}
If a computation region is described with $k$ codelet types then the overall cost of the codelets for that region is $C = \sum_{l=0}^{k} c_l$. Figure \ref{fig:cost_correlation} shows the correlation between the codelet cost model and the execution time of codelets. We use a set of codelets with different sizes from SpMV and SpTRSV  on different computation regions of the sparse matrices in Suitesparse~\cite{Davis2011}.  The X-axis shows the cost of each codelet,  and the Y-axis shows their corresponding execution time in seconds. As shown, the cost model predicts the performance of the codelets with a correlation of $(r^2=0.99)$.

\subsection{Differentiation Based PSC Detection}
\label{sec:dffpsc}
In this section, we explain how the \textit{first order partial difference} (FOPD) of the access functions in a computation region can be used to detect a codelet type. 



\begin{figure}
     \centering
     \includegraphics[width=0.88\linewidth]{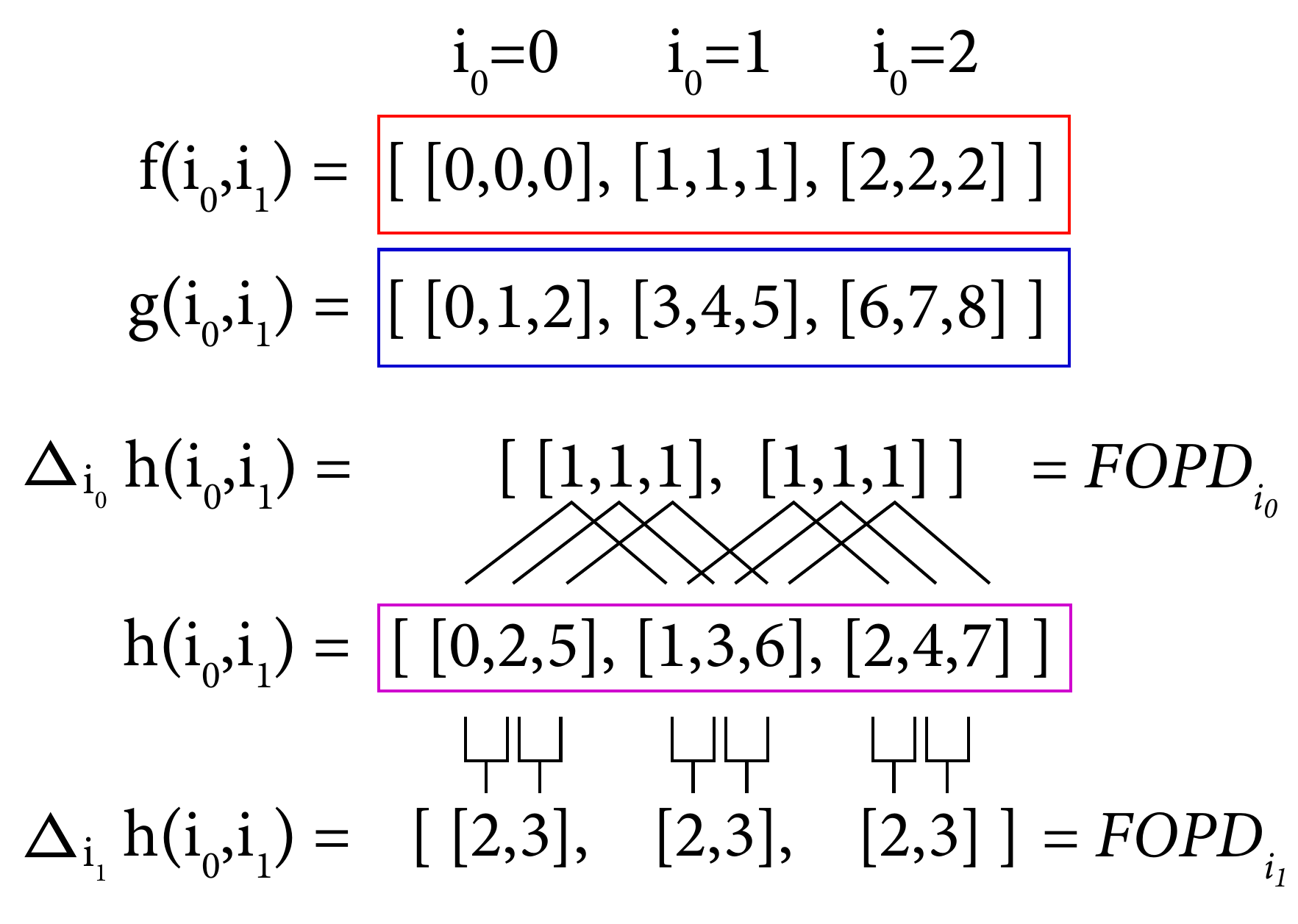}
     \caption{Illustrates the process of taking the derivative of the access function $h$ with respect to $i_0$ and $i_1$.}
     \label{fig:partial_derivative_example}
 \end{figure}

\textit{First Order Partial Differentiation (FOPD$_i$).} Given the data access function $f$ with iteration space of $\mathcal{I}=[i_0,i_1]$, the first order partial difference of $f$ with respect to $i_1\in\mathcal{I}$ is computed as $\frac{\Delta }{\Delta i_1}f(\mathcal{I}) = \Delta_{i_{1}} f =f(i_0,i_1+1)-f(i_0,i_1)$. FOPD shows if accesses to a data space are strided with respect to the induction variable $i_1$. Figure~\ref{fig:partial_derivative_example} illustrates the process of computing the FOPD for the access function $h$ given the computation region shown in Figure~\ref{fig:dataspace_example}. For example, the FOPD of $h$ evaluated at $i_0=1, i_1=1$ is $\Delta_{i_{1}} h(1,1) = 3$.

 FOPD of access functions are used to distinguish types of codelets by finding strided access functions. Given a codelet with three access functions and the iteration space of $\mathcal{I}=[i_0, i_1]$, 
 an access function $f$ is strided if its FOPDs with respect to $\mathcal{I}$ are equal in the entire iteration space. In other words, $f$ is strided if  the elements in $\Delta_{i_0} f(i_0,i_1)$ are equal to each other and similarly for elements in $\Delta_{i_1} f(i_0,i_1)$. %
 With the strided definition above, all codelet types can be defined per definitions in Section~\ref{sec:pscclas}. 
%
%
For example, the first three accesses  in function $g(i_0, i_1)$ in Figure~\ref{fig:partial_derivative_example} are to consecutive locations (0, 1, 2) in \texttt{Ax}. The FOPD of the first two accesses, wrt. $i_1$, is FOPD$_{i_1}(0,0):g(0,1)-g(0,0)=1$ and for the second and third accesses  is FOPD$_{i_1}(0,1):g(0,2)-g(0,1)=1$. Similarly the FOPD of the accesses for $i_0$ are FOPD$_{i_0}(0,0):g(1,0)-g(0,0)=3$ and FOPD$_{i_0}(1,0):g(2,0)-g(1,0)=3$.
Since FOPD$_{i_1}(0,0)$ and FOPD$_{i_1}(0,1)$ are equal, and FOPD$_{i_0}(0,0)$ and FOPD$_{i_0}(1,0)$ are equal, the function in the iteration space  $\mathcal{I}=\{i=0 \land 0\leq j \leq3\}$ is strided. Because function $f$ is also strided and $h$ is not strided, the codelet is categorized as a PSC type I. 


\begin{lstlisting}[label={lst:input},mathescape=true,numbers=none,keywords={Kernel, Pattern, int, return},numbers=left,
caption={The input code to \DDT{}.}]
#include "DDF.h"
int main(){
 Kernel SpMV;
 SpMV.generate_c("spmv.h", Arch::x86);
 return 1;
}
\end{lstlisting}

\begin{lstlisting}[label={lst:driver},mathescape=true,numbers=none,keywords={Kernel, Pattern, int, return},numbers=left,
caption={A sample driver code to call the \DDT{}-generated code.}]
#include "DDF.h"
#include "spmv.h"
int main(){
  Matrix A("input.mtx");
  Vector x("x.mtx"), y("y.mtx");
 /// ------ Inspector ------ ///
 vector<codelet *> clist=DDF_Inspector(A.pattern(), SpMV);
 /// ------ Executor ------ ///
 SpMV_Vectorized(clist, A, x, y);
 return 1;
}
\end{lstlisting}

\section{\DDT{}: Differentiating Data Access Functions to Mine PSC Codelets}



\DDT{} is an inspector-executor framework that finds partially strided codelets from an input code and the matrix sparsity pattern  using an inspector and generates a vectorized code as its executor. 
%
The \DDT{} input specification is shown in Listing~\ref{lst:input}. It takes the input kernel, SpMV or SpTRSV, and the target architecture type and generates a driver code shown in Listing~\ref{lst:driver} and a vectorized code in \texttt{spmv.h}.  The vectorized code is generated when \texttt{generate\_c} is called in line 4 of Listing~\ref{lst:input}. 
In lines 4--5 of the Listing~\ref{lst:driver}, the input matrix $A$ and vectors $x$ and $y$ are loaded. \DDT{}'s inspector then creates a list of codelets in line 7, and the vectorized code executes in line 9. 

\subsection{The \DDT{} Inspector}
\DDT{}'s inspector uses Algorithm \ref{alg:mark_adjacent} to generate efficient code for the input sparse kernel. The algorithm first creates groups of independent iterations and then for the computation regions in each group finds a best codelet combination to generate. 

\subsubsection*{Inputs and output} The inputs to the \DDT{} inspector algorithm are a kernel code $\mathcal{K}$, i.e., SpMV or SpTRSV, and the pattern of the input matrix $\mathcal{P}$. The algorithm generates a list of codelets $cList$ to be used in the vectorized code. 
In \DDT{}, the kernel code $\mathcal{K}$ is represented with an abstract syntax tree (AST) that represents loops and operations.
The pattern $\mathcal{P}$ is stored in a compressed sparse row (CSR) storage. 

\begin{algorithm}[t]
\begin{small}
\caption{\label{alg:mark_adjacent} \DDT{} Inspector}
\SetKwInOut{Input}{Input}
\SetKwInOut{Output}{Output}
\SetKwRepeat{Do}{do}{while}
\Input{$\mathcal{K},\mathcal{P}$}
\Output{$cList$}
\tcc{1) Grouping independent memory accesses}
$f_0, f_1, f_2, \mathcal{I} \leftarrow computeAccessFunctions(\mathcal{K},\mathcal{P})$\;\label{lin:acc1}
$\mathcal{G} \leftarrow findDependencies(f_0, f_1, f_2, \mathcal{I})$\;\label{lin:dep1}
$\mathcal{S} \leftarrow partitionIterationSpace(\mathcal{G}, \mathcal{I})$\;\label{lin:par1}
\tcc{2) Codelet Creation}
\For{$p \in \mathcal{S}$}{\label{lin:loopb2}
FOPDs $\leftarrow ComputeFOPD(p,f_0,f_1,f_2)$\;\label{lin:diffp}
$\mathcal{R} \leftarrow GetConsecutiveIterations(p)$\;\label{lin:ovr}
\For{$r \in \mathcal{R}$}{\label{lin:inloopb}
  $clb \leftarrow BLAS\_first(r ,$ FOPDs)\;\label{lin:p1}
  $clp1 \leftarrow PSCI\_first(r ,$ FOPDs)\;\label{lin:p2}
  $clp2 \leftarrow PSCII\_first(r ,$ FOPDs)\;\label{lin:p3}
  $min\_cl \leftarrow min\_cost(clb, clp1, clp2)$\;\label{lin:p4}
  $cList.append(min\_cl)$\;\label{lin:append}
 }\label{lin:inloope}
}\label{lin:loope2}
\end{small}
\end{algorithm}

\subsubsection{Grouping independent memory accesses}
The first step of the inspector algorithm computes memory access functions of the input code and then generates  a graph $\mathcal{G}$ and uses it to  create groups of operations that are independent to support parallelism. 
Function $computeAccessFunctions$ in line~\ref{lin:acc1} of Algorithm~\ref{alg:mark_adjacent} uses kernel $\mathcal{K}$ and goes over $\mathcal{P}$ to compute access functions and the iteration space of the input kernel. Then $findDependencies$ in line~\ref{lin:dep1} uses access functions to compute a graph $\mathcal{G}$. The vertices in this graph represent iterations of the outermost loop $i\in\mathcal{I}=[i,j]$ and its edges are the flow dependencies across iterations of $i$. 
In kernels with loop carried dependencies, $\mathcal{G}$  is passed to the LBC algorithm~\cite{Cheshmi2019} to create independent parallel workloads. For sparse kernels with parallel loops, $\mathcal{G}$ does not have any edges, thus instead, the function groups iterations based on the number of operations. Finally, iterations of the outermost loop are grouped and the groups are stored in $\mathcal{S}$ using $partitionIterationSpace$ in line 3.

\subsubsection{Codelet creation} 

For every group of iterations $p\in\mathcal{S}$, the algorithm generates the FOPD of the access functions in line 5. Function $GetConsecutiveIterations$ in line~\ref{lin:ovr} groups operations across $t$ consecutive iterations and creates a number of computation regions stored in $\mathcal{R}$. 
To reduce the overhead of mining, the algorithm limits the search window in line 6 to $t$ consecutive iterations. Our experiments show that this does not lead to slowdowns because consecutive iterations in sparse codes typically operate on consecutive rows of a matrix, and thus their vectorization can potentially improve spatial locality. 

\begin{figure}
    \centering
    \includegraphics[width=0.9\linewidth]{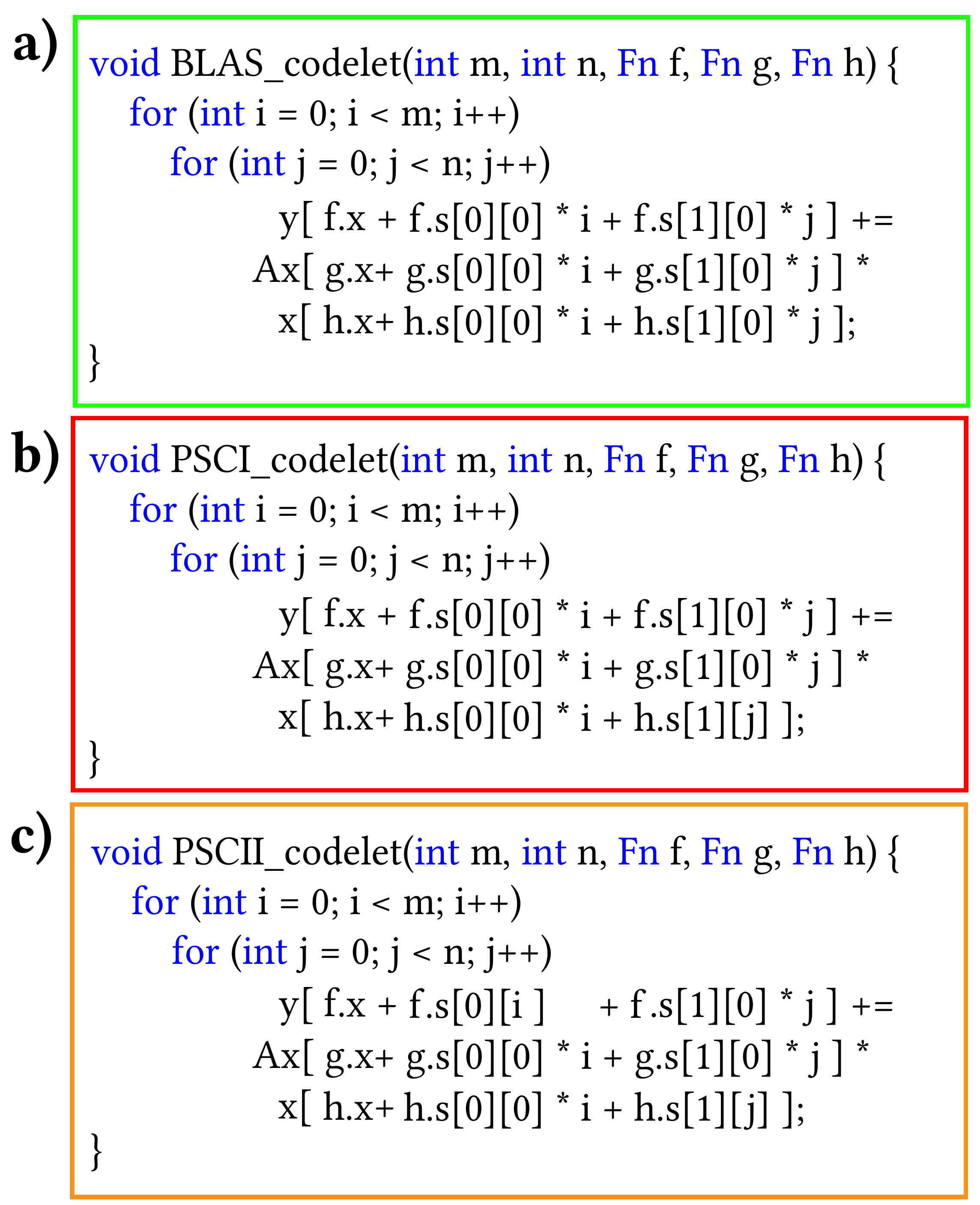}
    \caption{Shows generic codes used to evaluate different codelet types; all codelets are parameterized by m, n which bound the outer and inner dimensions respectively. Each type also uses f,g and h which are special structs containing offset information used to process each codelet.}
    \label{fig:generic_codelets}
\end{figure}

Lines 8-12 in the algorithm  inspect each computation region $r$ and finds the best codelet combination for that region using three different strategies, i.e.  $BLAS\_first$ ,  $PSCI\_first$, and $PSCII\_first$. Since the strategies mine for codelets, they take as input the FOPDs computed from line 5 along with $r$. The BLAS-first strategy prioritizes finding BLAS codelets and then mines for PSCs. The PSC I-first strategy mines for PSC type I codelets first, and the PSC II-strategy looks for PSC type II codelets and then other codelets.  To reduce the execution time of each strategy,  we store the list of codelets for each mined region and look up this list when executing the other strategies. The costs of codelets from each approach are stored in constants $clb$, $clp1$, and $clp2$ and then in line 11 the combination with the lowest cost is chosen. Line 12 appends the most efficient  codelet combination found for the current computation region to the output list of codelets in the Algorithm, i.e. $cList$.

\subsection{\DDT{} Executor}

The \DDT{} driver  runs the executor code, as shown in line 9 of listing 2. To 
efficiently use the instruction cache and to eliminate the need for recomputing a new executor code per matrix pattern, \DDT{} generates a parametric code.  
 %
%
%
%
%
%
The parametric code for the three types of codelets, BLAS, PSC I, and PSC II are shown in Figure~\ref{fig:generic_codelets}. 
As shown each generic code of a codelet class takes the data access functions as input and can vectorize any codelet of the same class.
These generic codelets are called inside a  switch statement in the executor code, shown in  Listing~\ref{lst:vecc}. In the executor, each mined codelet in the codelet list is mapped to a generic codelet using this switch statement. 
A change in the input sparsity pattern results in a different codelet list, however, the same parametric code is still used.   
For matrices of different size, the codelets list size changes, however, the same executor code in  Listing~\ref{lst:vecc} can be used. Thus, the code size is invariant to the size of matrix.

\begin{lstlisting}[label={lst:vecc},mathescape=true,keywords={Kernel, Pattern, switch, case, for, int, double, void, break},numbers=left,float,
caption={The vectorized code for SpMV generated by \DDT{} in \texttt{spmv.h}.}]
#include "Codeletsx86.h"

void vectorized_code(vector<Codlet*> clsit, CSR* A, double *x, double *y){
 for(int i = 0; i < clist.partitions(); i++){
#pragma omp parallel 
  for(int j = 0; j < clist[i].size(); j++){
   switch(clist[i][j].type){
    case BLAS:
     BLAS_codelet();
     break;
    case PSCI:
     PSCI_codelet();
     break;
    case PSCII:
     PSCII_codelet();
     break;
}}}}

\end{lstlisting}

\section{Results}
We evaluate the performance of \DDT{} using two kernels, sparse triangular solver (SpTRSV) and sparse matrix-vector multiplication (SpMV).  \DDT{} is compared to two other inspector-executor approaches, i.e. Sympiler and PIC~\cite{Augustine2019}. 
We also compare \DDT{} to libraries  CSR5~\cite{liu2015csr5}, MKL~\cite{mkl_lib}, and our implementation of SpMV using the ELLPACK~\cite{saad2003iterative} storage format (referred to as ELL in the figures). Sympiler does not support SpMV, and SpTRV is not supported by the CSR5, PIC, and ELL implementations. Hence those tools are omitted from the respective figures.

%



The set of symmetric positive definite (SPD) matrices is used for evaluation.
These matrices are selected from the SPD symmetric matrices in the SuiteSparse repository~\cite{davis2011university} and are diverse in both size and sparsity pattern. For the SpTRSV evaluations, we only use the lower triangular half of the SPD symmetric matrices. Matrices with ID 0=30 are the thirty largest matrices in the repository, Matrix IDs 31-49 are L-factors~\cite{davis2005cholmod}  of the largest matrices, and  10 small matrices  (ID = 50-60) with 50-350K nonzeros are also included. L-factors matrices are included to compare the performance of \DDT{}
 to Sympiler because Sympiler is best suited to optimize these matrices. Small matrices are included to compare with PIC as it does not scale to larger problems. The test-bed architectures are an Intel(R) Xeon(R) Gold 5115 CPU (2.8GHz, 14080K L3 Cache) with 20 cores and 64GB of main memory (Intel) and an AMD Ryzen 3900x CPU (3.8GHz, 64MB L3 Cache) with 12 cores and 32GB of main memory (AMD). The performance of kernels is reported on both architectures but the analysis is only shown for the Intel machine. 
 All generated code, implementations of different approaches, and library drivers are compiled with GCC v.7.2.0 compiler and with the \texttt{-O3} flag. All benchmarks are executed 5 times and the median value of all runs is reported.
 To report the performance in GFLOP/s, we compute the theoretical floating-point operations for each kernel and matrix and divide it by execution times of each tool. 
 
The sequential implementation of SpMV CSR and SpTRSV CSR are used as a baseline. 
Since the code for PIC is not publicly available, we created an in-house inspector-executor implementation of their approach with feedback from the authors of \cite{Augustine2019}. 
PIC was originally developed for a single thread, however, we extended its support to be parallel and report the best performance between the two implementations in all figures. A timeout of 4 hours was used for all runs (including the inspector and the executor time). Thus, PIC does not have data points in respective figures for large matrices. 
Also, our implementation of ELL which is based on \cite{saad2003iterative}, has missing points in some figures because the number of non-zero fill-ins exceeds  the main memory of our test-bed architecture. 

\subsection{SpMV Performance}
\begin{figure*}[t]
    \centering
    \includegraphics[width=0.93\linewidth]{{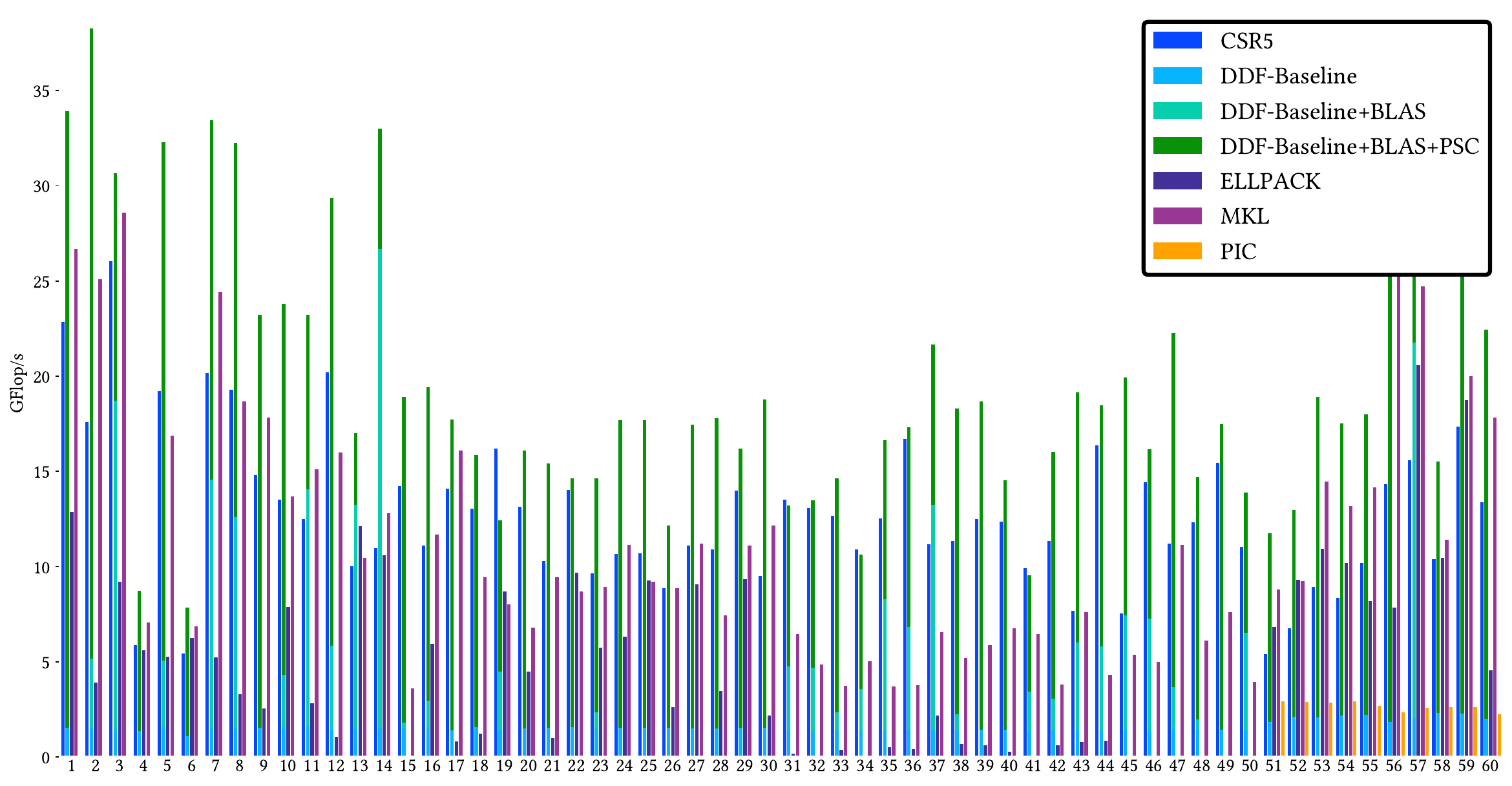}}
    \includegraphics[width=0.93\linewidth]{{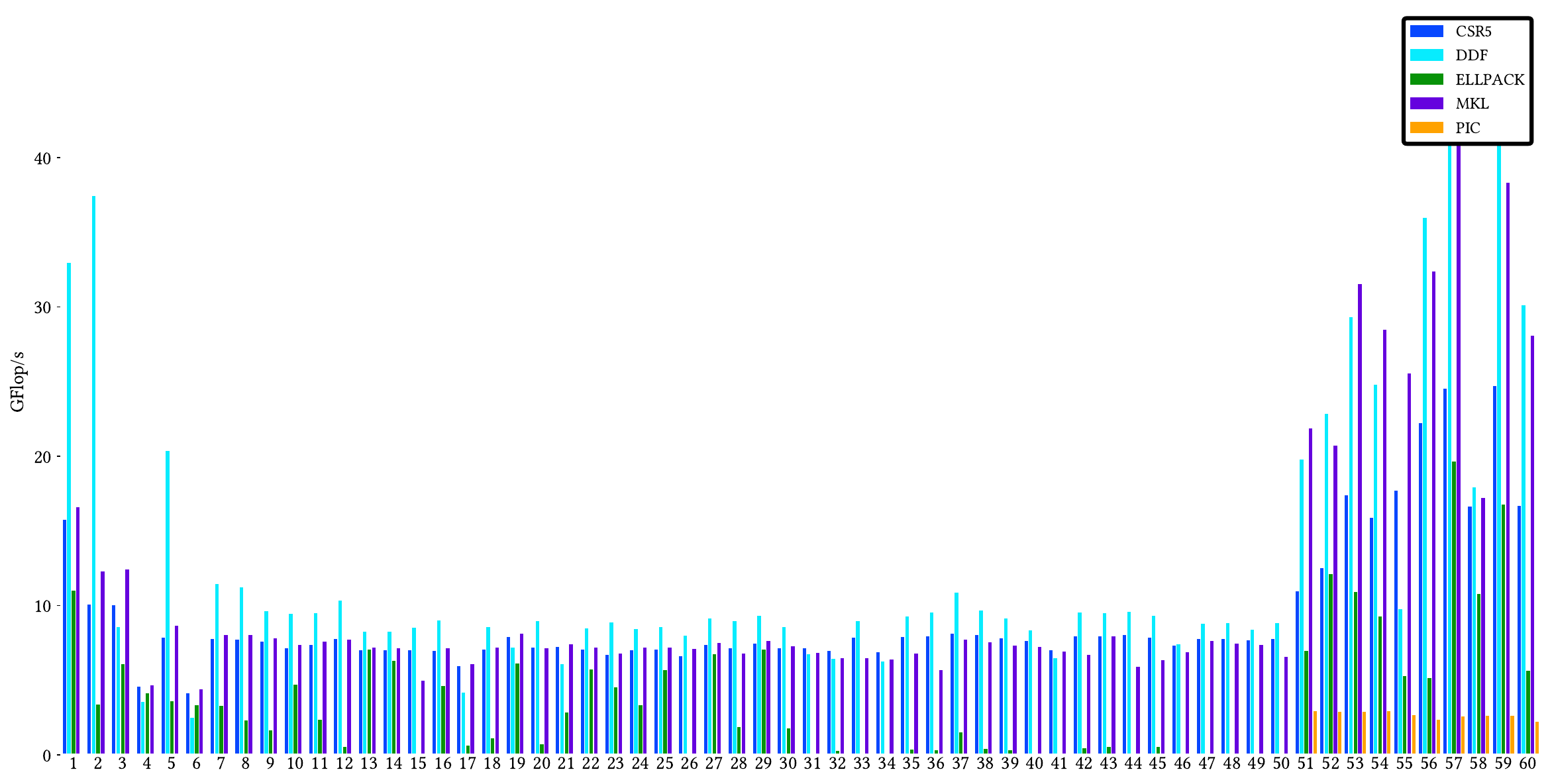}}
    \caption{\textit{Top}: Shows \DDT{} speedups for the SpMV kernel over the single threaded baseline SpMV CSR algorithm ran on an Intel architecture. Speedups also reported for parallel MKL, parallel CSR5, parallel ELLPACK and single threaded PIC \cite{Augustine2019} over the single threaded baseline SpMV CSR algorithm. \textit{Bottom}: Shows \DDT, MKL, CSR5, ELLPACK and PIC speedups for the SpMV kernel using an AMD architecture.}
    \label{fig:spmv_speedup}
\end{figure*}

The performance of the SpMV kernel for \DDT{}, MKL, CSR5, ELLPACK, baseline, and PIC is shown in Figure \ref{fig:spmv_speedup}. As shown, \DDT{} is faster than all other tools for over 88\% of the matrices tested. 
\DDT{}'s code is on average 1.93x, 1.42x, 10.57x, and 7.19x faster than in order MKL, CSR5, ELLPACK, and PIC respectively on the Intel architecture. \DDT{}'s code is on average 1.22x, 1.29x, 8.54x, and 10.93x faster than in order MKL, CSR5, ELLPACK, and PIC respectively on the AMD architecture.

\begin{figure}[t]
    \centering
    \includegraphics[width=\linewidth]{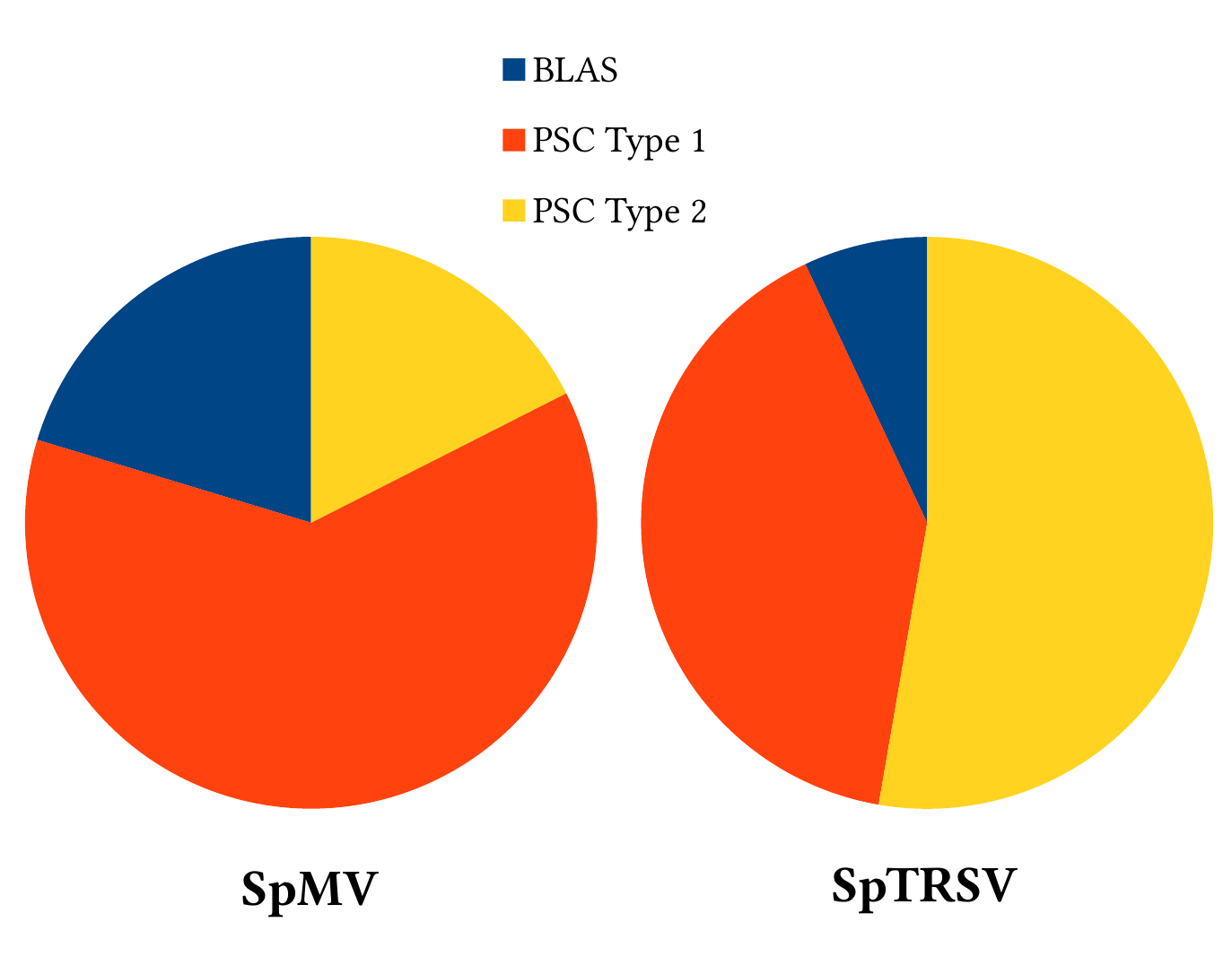}
    \caption{Shows percentage of points for all matrices processed by BLAS, PSC type 1 or PSC type 2 codelets for SpMV and SpTRSV.}
    \label{fig:codelet_breakdown_piechart}
\end{figure}

 To demonstrate the effect of partially strided codelets on the performance of  \DDT{}, in Figure~\ref{fig:codelet_breakdown_piechart} we use a stacked bar for \DDT{}. The stacks show the GFlop/s for the baseline code (input to \DDT{}), for  \DDT{} when it only mines for BLAS codelets   (\DDT-baseline+BLAS and refer to \DDT{} BLAS\_only), and from running the entire \DDT{} algorithm, i.e. Algorithm 1, that mines for BLAS and PSC (shown with \DDT{} or \DDT-baseline+BLAS+PSC in the figure).  
 As shown, \DDT{} is on average 10$\times$ faster than \DDT{} BLAS\_only which demonstrates the importance of using PSC codelets. 
Figure \ref{fig:codelet_breakdown_piechart} shows  the percentage of operations that are vectorized in the generated code from \DDT{} with PSC I, PSC II, and BLAS codelets, obtained by averaging over all matrices in the benchmark. As shown, over 80\% of the operations in SpMV are  vectorized with PSC codelets.  

\begin{figure}[t]
    \centering
    \includegraphics[width=0.8\linewidth]{{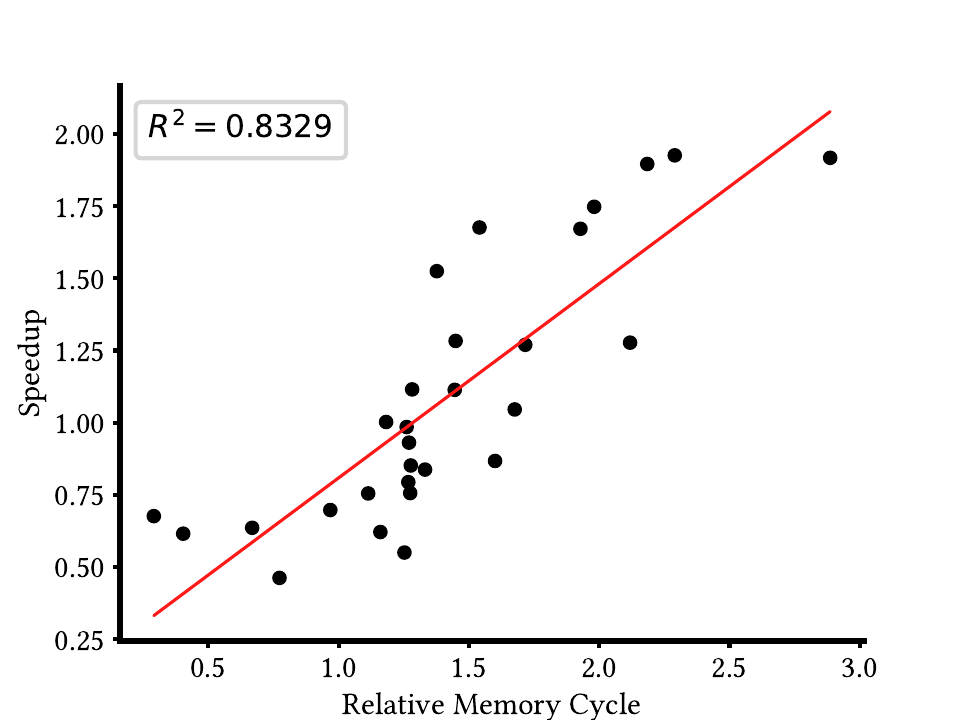}}
    \caption{Shows the correlation between speedup over the baseline while profiling the code with PAPI ~\cite{terpstra2010collecting} and the relative memory cycle for matrices used in the test set for SpMV. Relative memory cycle is the average memory cycle of the baseline over the average memory cycle of the DDF framework codes.}
    \label{fig:average_memory_cycle_comparison}
\end{figure}

The PSC codelets  improve data locality in \DDT{}'s generated code. 
Figure \ref{fig:average_memory_cycle_comparison} shows the relation between \DDT{}'s  performance and the performance of the parallel (OpenMP) version of the baseline code. 
Average memory access latency~\cite{hennessy_quantitative} is used as a measure for locality and is computed by gathering the number of misses and accesses to L1, L2, and LLC caches using the PAPI~\cite{terpstra2010collecting} performance counters.   
Figure~\ref{fig:average_memory_cycle_comparison} shows the coefficient of determination or R$^2$ is 0.83 which indicates a good correlation between speedup and the memory access latency. 


To further explain why \DDT{} is faster than 
other tools, we conduct a few experiments and report the resulting average over all matrices in this paragraph. 
For MKL, we compute its average memory access latency, which is  4.36x slower than that of \DDT{}, contributing to the worse performance compared to \DDT{}. To compare to CSR5, we count the number of instructions. CSR5 executes 1.73x more instructions compared to \DDT{}. This is potentially due to the overhead of the segmented sum calculations used in their approach to improving vectorization and load balance.  \DDT{} is on average  10.57x faster than ELL, because the percentage of fill-in in the ELL implementation is 800\% relative to the non-zero elements in the matrix, significantly increasing the number of operations in their method. To conclude, the SpMV code of \DDT{} is faster than existing implementations because it improves data locality and/or reduces the number of instructions via vectorization. 



\subsection{SpTRSV Performance}

\begin{figure*}[t]
    \centering
    \includegraphics[width=0.78\textwidth]{{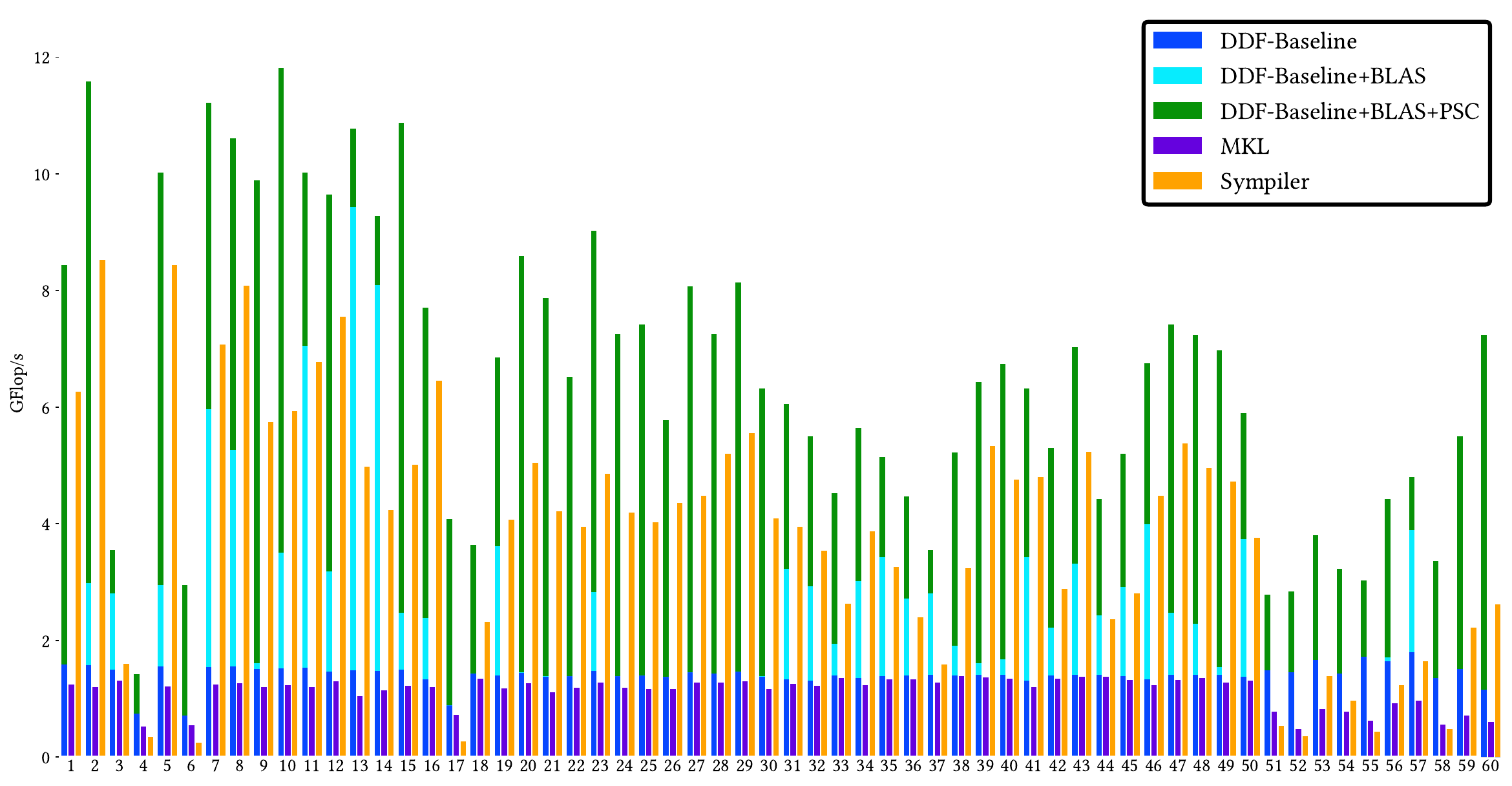}}
    \includegraphics[width=0.78\textwidth]{{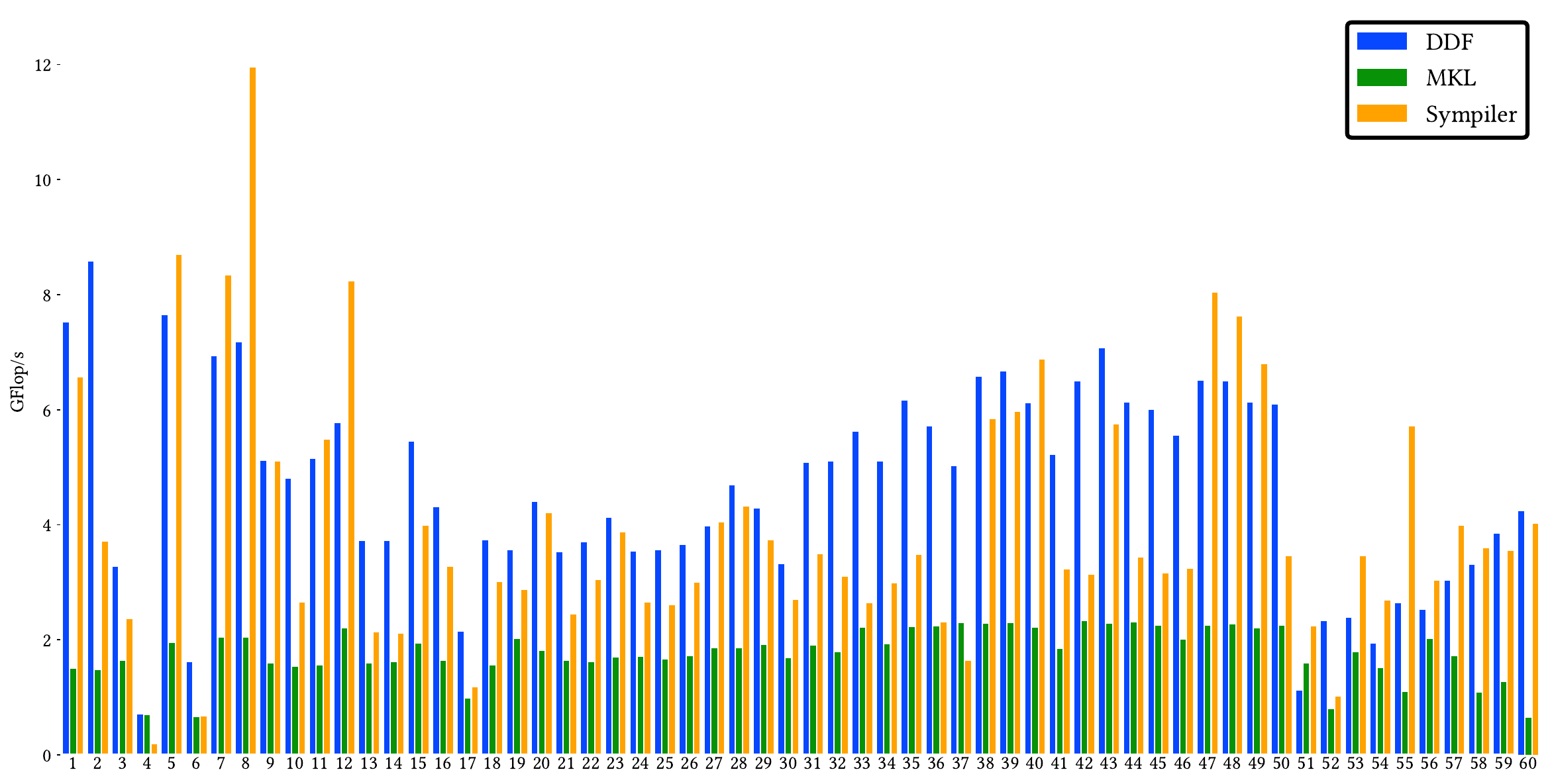}}
    \caption{\textit{Top}: Shows breakdowns for \DDT{} speedups, when enabling BLAS and PSC codelet mining, MKL and Sympiler speedups over the single threaded baseline SpTRSV CSR algorithm on Intel archtecture. \textit{Bottom}: Shows \DDT{}, MKL and Sympiler speedups on AMD architecture.}
    \label{fig:sptrsv_speedup}
\end{figure*}

Figure \ref{fig:sptrsv_speedup} compares the  performance of SpTRSV  using \DDT{}, MKL, and Sympiler. The stacked bar of \DDT{} shows the additional performance gained from mining partially strided codelets; the trend is similar to that in Figure \ref{fig:spmv_speedup} for SpMV. On average \DDT{} is faster than MKL, Sympiler, and the baseline  4.5$\times$, 1.79$\times$, and 3.62$\times$ respectively for the intel architecture. On the AMD architecture, \DDT{} is faster than MKL and Sympiler 2.66, 1.38$\times$ respectively.

As shown in Figure~\ref{fig:sptrsv_speedup}, partially strided codelets are the main contributors to the overall performance of the \DDT{}'s SpTRSV code. On average, \DDT{} is 3.9$\times$ faster compared to when only BLAS codelets are generated.
Similar to SpMV, PSCs contribute to optimizing 78\% of the operations over all matrices (Figure 6b). However, the number of PSC II codelets has increased from 18\% in SpMV to 53\% in SpTRSV. The number of computational regions with more than one strided access function is small, due to the existing dependencies in the SpTRSV kernel, thus, more PSC type II codelets are generated. 
%
Similar to the SpMV kernel, the mined PSCs in \DDT{} improve locality. 
The correlation coefficient between the speedup and relative memory cycle is 0.67 which is consistent with the trend  in SpMV.

While MKL provides an efficient and vectorized implementation for single-threaded SPTRV executions, it's not optimized to execute on parallel processors; the performance of MKL's parallel code is similar to its serial implementation. 
Sympiler performs well for matrices that contain row-blocks or can be padded with up to 30\% nonzeros to create row-blocks. These row-blocks are converted to BLAS calls and thus improve locality. 
 L-factors are generated from a Cholesky factorization of the original matrices in
the SuiteSparse collection. Since L-factors are created with a supernodal factorization process \cite{chen2008algorithm}, the resulting matrices are well-structured and have many row-blocks.  As shown, Sympiler's performance on L-factor matrices (IDs 31-50)  is close to that of \DDT{}'s with an average speedup of 1.18$\times$, while for all other matrices \DDT's performs on average time 2.1$\times$ better than Sympiler.

\subsection{The Inspector Overhead}
We compare the inspection time of \DDT{} to that of other  inspector-executors/code-generation frameworks, i.e. Sympiler and PIC. We compute the number of executor runs (NER) that amortize the cost of the inspector using \\$ \frac{ \>\> Inspector \>\> Time}{Baseline\> Time -  \>\> Executor \>\> Time} $. The \textit{baseline} time is obtained by running sequential implementation of the kernel. 
The PIC inspector would timeout for over 83.3\% of matrices because of its large inspection overhead and code compilation time. For small matrices that PIC would execute, more than one million executor runs are needed to amortize the cost of the inspection. \DDT{}'s inspection time is on average 0.5 seconds with an average NER of 15 for SpMV. The inspection time of \DDT{} and Sympiler are similar with an average  NER of less than 100 for both tools for SPTRV. The largest matrices in our benchmark are inspected in less than 7 seconds with \DDT{}. Sparse kernels such as SpMV and SpTRSV are typically used in iterative solvers, for example, to compute a residual in each iteration or to apply a preconditioner per iteration. Even with preconditioning, these solvers typically converge to a solution after tens of thousands of iterations \cite{benzi2000robust,kershaw1978incomplete,papadrakakis1993accuracy}  and hence inspector-executor frameworks such as \DDT{} and Sympiler lead to noticeable speedups as their inspection time overheads are amortized after a few initial iterations of the solver. 

\section{Related Work}

Numerous hand-optimized libraries \cite{intel-alt,eigenweb} and implementations \cite{Li2003,Hutchinson1995,davis2005cholmod,davis2004algorithm} exist that optimize the performance of sparse matrix computations for different parallel architectures and also optimize vectorization on a single core. Libraries such as MKL and Eigen as well as implementations in \cite{Yesil2020,li2013gpu,williams2007optimization,kamin2014optimization,merrill2016merge} optimize the performance of SpMV on shared memory architectures and improve SIMD vectorizability. A number of library implementations such \cite{saad2003iterative,vuduc2003automatic, vuduc2005fast,li2020vbsf,liu2015csr5,tang2015optimizing,xie2018cvr} reorganize data and computation to increase opportunities for vectorization. A class of these libraries implement and optimize sparse kernels based on available storage formats; for example \cite{kreutzer2014unified} optimizes SpMV based on ELLPACK.
Other libraries work best for matrices arising from specific applications such as \cite{Yesil2020}, which optimizes SpMV for
large matrices from graph analytics, or KLU \cite{davis2010algorithm} which works best for circuit simulation problems.


Domain-specific compilers use domain information to enable the application of code transformations and optimizations such as vectorization. For example domain-specific compilers such as \cite{jones1993partial,quillere2000generation,Spampinato2014} optimize signal processing applications, stencil computations are optimized in \cite{jones1993partial,quillere2000generation,spampinato2014basic}, and matrix assembly and mesh analysis is optimized in \cite{alnaes2014unified,luporini2017algorithm}. While these compilers provide highly-optimized code for the applications that they accelerate, they do not generate specialized code for the input pattern and the computations   do not have indirect accesses.

\textit{Inspector-Executor approaches} inspect the irregular access patterns of sparse matrix computations at run-time to enable the automatic optimization of sparse codes \cite{Shin2010,Tang2011,VanDerSpek2011,vasilache2006polyhedral,Venkat2015,agrawal1995interprocedural,das1995index,Rodriguez,saltz1990run}. The index array accesses of the sparse code is analyzed using an inspector and the information is used at run-time to execute the code efficiently. The sparse polyhedral framework \cite{lamielle2010enabling,strout2012set,venkat2016automating} uses uninterpreted function symbols to express regular and irregular segments of sparse codes. As a result it is able to automatically generate inspector executors at compile time that can resolve data dependencies in sparse computations. These approaches do not generate code that is specialized for the sparsity of the input matrix. Sympiler \cite{Cheshmi2017} and ParSy \cite{Cheshmi2019} are amongst the inspector executor frameworks that inspect the matrix sparsity pattern and as a result generate vectorized and parallel code specialized for the input sparsity. Their optimizations for tiling and vactorization are  based on detecting row-blocks that primarily exist in matrices  obtained from numerical factorizations. 

Augustine et al. \cite{Augustine2019} proposed an approach based on the Trace Reconstruction Engine \cite{Rodriguez2016,Ketterlin2008} where polyhedral representations are built by inspecting the sequence of addresses being accessed in the sparse matrix vector multiplication. A followup to this work, proposes to use program guided optimization for better vectorization \cite{Selva2019}. These approaches lead to generating code that is specialized for the sparsity pattern of the input matrix and improves SIMD vectorization in SpMV. However, their work can only support small matrices (below 0.5M nonzeros) because of inspector overheads and because the size of the generated code increases with the matrix size. Sparse computations with loop-carried dependencies such as the sparse triangular solver are also not supported. \DDT{} inspects memory address accesses to find SIMD vectorizable code for sparse triangular solve and SpMV and supports both small and large matrices.

\section{Conclusion}
In this work, we present partially strided codelets that enable the vectorization of computation regions with unstrided memory accesses in sparse matrix codes. We demonstrate how these codelets increase opportunities for vectorization in sparse codes and also improve data locality in their computation. A novel inspector-executor framework called \DDT{} is proposed. \DDT{} uses an efficient inspector to mine for PSCs with a memory access differentiation approach and as a result, generates highly efficient code for sparse kernels. The performance of the \DDT{}-generated code is compared to state-of-the-art library implementations and other inspector-executor frameworks for the sparse matrix-vector multiply and the sparse triangular solve kernels. We also demonstrate that the inspection overhead of \DDT{} is negligible.


\section{Acknowledgments}
This work was supported in part by NSERC Discovery Grants (RGPIN-06516, DGECR00303), the Canada Research Chairs program, and U.S. NSF awards NSF CCF-1814888, NSF CCF-1657175; used the Extreme Science and Engineering Discovery Environment (XSEDE) [Towns et al. 2014] which is supported by NSF grant number ACI-1548562; and was enabled in part by Compute Canada and Scinet\footnote{www.computecanada.ca}

\bibliographystyle{IEEEtran}
\bibliography{refs}

\begin{thebibliography}{10}
\providecommand{\url}[1]{#1}
\csname url@samestyle\endcsname
\providecommand{\newblock}{\relax}
\providecommand{\bibinfo}[2]{#2}
\providecommand{\BIBentrySTDinterwordspacing}{\spaceskip=0pt\relax}
\providecommand{\BIBentryALTinterwordstretchfactor}{4}
\providecommand{\BIBentryALTinterwordspacing}{\spaceskip=\fontdimen2\font plus
\BIBentryALTinterwordstretchfactor\fontdimen3\font minus
  \fontdimen4\font\relax}
\providecommand{\BIBforeignlanguage}[2]{{%
\expandafter\ifx\csname l@#1\endcsname\relax
\typeout{** WARNING: IEEEtran.bst: No hyphenation pattern has been}%
\typeout{** loaded for the language `#1'. Using the pattern for}%
\typeout{** the default language instead.}%
\else
\language=\csname l@#1\endcsname
\fi
#2}}
\providecommand{\BIBdecl}{\relax}
\BIBdecl

\bibitem{Cheshmi2017}
\BIBentryALTinterwordspacing
K.~Cheshmi, S.~Kamil, M.~M. Strout, and M.~M. Dehnavi, ``{Sympiler:
  Transforming sparse matrix codes by decoupling symbolic analysis},'' in
  \emph{Proceedings of the International Conference for High Performance
  Computing, Networking, Storage and Analysis, SC 2017}.\hskip 1em plus 0.5em
  minus 0.4em\relax New York, NY, USA: Association for Computing Machinery,
  Inc, nov 2017, pp. 1--13. [Online]. Available:
  \url{https://dl.acm.org/doi/10.1145/3126908.3126936}
\BIBentrySTDinterwordspacing

\bibitem{Augustine2019}
\BIBentryALTinterwordspacing
T.~Augustine, L.~N. Pouchet, J.~Sarma, and G.~Rodr{\'{i}}guez, ``{Generating
  piecewise-regular code from irregular structures},'' in \emph{Proceedings of
  the ACM SIGPLAN Conference on Programming Language Design and Implementation
  (PLDI)}.\hskip 1em plus 0.5em minus 0.4em\relax New York, NY, USA:
  Association for Computing Machinery, jun 2019, pp. 625--639. [Online].
  Available: \url{https://dl.acm.org/doi/10.1145/3314221.3314615}
\BIBentrySTDinterwordspacing

\bibitem{bulucc2009parallel}
A.~Bulu{\c{c}}, J.~T. Fineman, M.~Frigo, J.~R. Gilbert, and C.~E. Leiserson,
  ``Parallel sparse matrix-vector and matrix-transpose-vector multiplication
  using compressed sparse blocks,'' in \emph{Proceedings of the twenty-first
  annual symposium on Parallelism in algorithms and architectures}, 2009, pp.
  233--244.

\bibitem{intel-alt}
\emph{Intel Math Kernel Library. Reference Manual}.\hskip 1em plus 0.5em minus
  0.4em\relax Intel Corporation, 2009, santa Clara, USA. ISBN 630813-054US.

\bibitem{Kreutzer2013}
\BIBentryALTinterwordspacing
M.~Kreutzer, G.~Hager, G.~Wellein, H.~Fehske, and A.~R. Bishop, ``{A unified
  sparse matrix data format for efficient general sparse matrix-vector multiply
  on modern processors with wide SIMD units},'' \emph{SIAM Journal on
  Scientific Computing}, vol.~36, no.~5, pp. C401--C423, jul 2013. [Online].
  Available: \url{http://arxiv.org/abs/1307.6209
  http://dx.doi.org/10.1137/130930352}
\BIBentrySTDinterwordspacing

\bibitem{Liu2013}
\BIBentryALTinterwordspacing
X.~Liu, M.~Smelyanskiy, E.~Chow, and P.~Dubey, ``{Efficient sparse
  matrix-vector multiplication on x86-based many-core processors},'' in
  \emph{Proceedings of the International Conference on Supercomputing}.\hskip
  1em plus 0.5em minus 0.4em\relax New York, New York, USA: ACM Press, 2013,
  pp. 273--282. [Online]. Available:
  \url{http://dl.acm.org/citation.cfm?doid=2464996.2465013}
\BIBentrySTDinterwordspacing

\bibitem{Liu2015}
\BIBentryALTinterwordspacing
W.~Liu and B.~Vinter, ``{CSR5: An Efficient Storage Format for Cross-Platform
  Sparse Matrix-Vector Multiplication},'' \emph{Proceedings of the
  International Conference on Supercomputing}, vol. 2015-June, pp. 339--350,
  mar 2015. [Online]. Available: \url{http://arxiv.org/abs/1503.05032}
\BIBentrySTDinterwordspacing

\bibitem{Tang2015}
W.~T. Tang, R.~Zhao, M.~Lu, Y.~Liang, H.~P. Huyng, X.~Li, and R.~S.~M. Goh,
  ``{Optimizing and auto-tuning scale-free sparse matrix-vector multiplication
  on Intel Xeon Phi},'' in \emph{Proceedings of the 2015 IEEE/ACM International
  Symposium on Code Generation and Optimization, CGO 2015}.\hskip 1em plus
  0.5em minus 0.4em\relax Institute of Electrical and Electronics Engineers
  Inc., mar 2015, pp. 136--145.

\bibitem{Chen2016}
\BIBentryALTinterwordspacing
L.~Chen, P.~Jiang, and G.~Agrawal, ``{Exploiting recent SIMD architectural
  advances for irregular applications},'' in \emph{Proceedings of the 14th
  International Symposium on Code Generation and Optimization, CGO 2016}.\hskip
  1em plus 0.5em minus 0.4em\relax New York, NY, USA: Association for Computing
  Machinery, Inc, feb 2016, pp. 47--58. [Online]. Available:
  \url{https://dl.acm.org/doi/10.1145/2854038.2854046}
\BIBentrySTDinterwordspacing

\bibitem{Xie2018}
\BIBentryALTinterwordspacing
B.~Xie, W.~Gao, J.~Zhan, Z.~Jia, L.~Zhang, X.~Liu, and X.~He, ``{CVR: Efficient
  Vectorization of SpMV on X86 Processors},'' in \emph{CGO 2018 - Proceedings
  of the 2018 International Symposium on Code Generation and Optimization},
  vol. 2018-February.\hskip 1em plus 0.5em minus 0.4em\relax New York, NY, USA:
  Association for Computing Machinery, Inc, feb 2018, pp. 149--162. [Online].
  Available: \url{https://dl.acm.org/doi/10.1145/3168818}
\BIBentrySTDinterwordspacing

\bibitem{liu2015csr5}
W.~Liu and B.~Vinter, ``Csr5: An efficient storage format for cross-platform
  sparse matrix-vector multiplication,'' in \emph{Proceedings of the 29th ACM
  on International Conference on Supercomputing}, 2015, pp. 339--350.

\bibitem{xie2018cvr}
B.~Xie, J.~Zhan, X.~Liu, W.~Gao, Z.~Jia, X.~He, and L.~Zhang, ``Cvr: Efficient
  vectorization of spmv on x86 processors,'' in \emph{Proceedings of the 2018
  International Symposium on Code Generation and Optimization}, 2018, pp.
  149--162.

\bibitem{saad2003iterative}
Y.~Saad, \emph{Iterative methods for sparse linear systems}.\hskip 1em plus
  0.5em minus 0.4em\relax SIAM, 2003.

\bibitem{Yesil2020}
S.~Yesil, A.~Heidarshenas, A.~Morrison, and J.~Torrellas, ``{Speeding up spmv
  for power-law graph analytics by enhancing locality vectorization},'' in
  \emph{International Conference for High Performance Computing, Networking,
  Storage and Analysis, SC}, vol. 2020-November.\hskip 1em plus 0.5em minus
  0.4em\relax IEEE Computer Society, nov 2020, pp. 1--15.

\bibitem{vazquez2010improving}
F.~Vazquez, G.~Ortega, J.-J. Fern{\'a}ndez, and E.~M. Garz{\'o}n, ``Improving
  the performance of the sparse matrix vector product with gpus,'' in
  \emph{2010 10th IEEE International Conference on Computer and Information
  Technology}.\hskip 1em plus 0.5em minus 0.4em\relax IEEE, 2010, pp.
  1146--1151.

\bibitem{Boulet1998}
P.~Boulet and P.~Feautrier, ``{Scanning polyhedra without Do-loops},'' in
  \emph{Parallel Architectures and Compilation Techniques - Conference
  Proceedings, PACT}.\hskip 1em plus 0.5em minus 0.4em\relax Institute of
  Electrical and Electronics Engineers Inc., 1998, pp. 4--11.

\bibitem{Chen2012}
\BIBentryALTinterwordspacing
C.~Chen, ``{Polyhedra scanning revisited},'' in \emph{Proceedings of the ACM
  SIGPLAN Conference on Programming Language Design and Implementation
  (PLDI)}.\hskip 1em plus 0.5em minus 0.4em\relax New York, New York, USA: ACM
  Press, 2012, pp. 499--508. [Online]. Available:
  \url{http://dl.acm.org/citation.cfm?doid=2254064.2254123}
\BIBentrySTDinterwordspacing

\bibitem{Kershaw1978}
D.~S. Kershaw, ``{The incomplete Cholesky-conjugate gradient method for the
  iterative solution of systems of linear equations},'' pp. 43--65, jan 1978.

\bibitem{Puschel2005}
\BIBentryALTinterwordspacing
M.~P{\"{u}}schel, J.~M. Moura, J.~R. Johnson, D.~Padua, M.~M. Veloso, B.~W.
  Singer, J.~Xiong, F.~Franchetti, A.~Ga{\v{c}}i{\'{c}}, Y.~Voronenko, K.~Chen,
  R.~W. Johnson, and N.~Rizzolo, ``{SPIRAL: Code generation for DSP
  transforms},'' in \emph{Proceedings of the IEEE}, vol.~93, no.~2.\hskip 1em
  plus 0.5em minus 0.4em\relax Institute of Electrical and Electronics
  Engineers Inc., 2005, pp. 232--273. [Online]. Available:
  \url{https://experts.illinois.edu/en/publications/spiral-code-generation-for-dsp-transforms}
\BIBentrySTDinterwordspacing

\bibitem{Spampinato2014}
\BIBentryALTinterwordspacing
D.~G. Spampinato and M.~P{\"{u}}schel, ``{A basic linear algebra compiler},''
  in \emph{Proceedings of the 12th ACM/IEEE International Symposium on Code
  Generation and Optimization, CGO 2014}.\hskip 1em plus 0.5em minus
  0.4em\relax New York, NY, USA: Association for Computing Machinery, feb 2014,
  pp. 23--32. [Online]. Available:
  \url{http://dl.acm.org/doi/10.1145/2544137.2544155}
\BIBentrySTDinterwordspacing

\bibitem{Tiwari2009}
A.~Tiwari, C.~Chen, J.~Chame, M.~Hall, and J.~K. Hollingsworth, ``{A scalable
  auto-tuning framework for compiler optimization},'' in \emph{IPDPS 2009 -
  Proceedings of the 2009 IEEE International Parallel and Distributed
  Processing Symposium}, 2009.

\bibitem{strout2012set}
M.~M. Strout, G.~Georg, and C.~Olschanowsky, ``Set and relation manipulation
  for the sparse polyhedral framework,'' in \emph{International Workshop on
  Languages and Compilers for Parallel Computing}.\hskip 1em plus 0.5em minus
  0.4em\relax Springer, 2012, pp. 61--75.

\bibitem{Venkat2015}
\BIBentryALTinterwordspacing
A.~Venkat, M.~Hall, and M.~Strout, ``{Loop and data transformations for sparse
  matrix code},'' in \emph{Proceedings of the ACM SIGPLAN Conference on
  Programming Language Design and Implementation (PLDI)}, vol. 2015-June.\hskip
  1em plus 0.5em minus 0.4em\relax New York, NY, USA: Association for Computing
  Machinery, jun 2015, pp. 521--532. [Online]. Available:
  \url{https://dl.acm.org/doi/10.1145/2737924.2738003}
\BIBentrySTDinterwordspacing

\bibitem{strout2018sparse}
M.~M. Strout, M.~Hall, and C.~Olschanowsky, ``The sparse polyhedral framework:
  Composing compiler-generated inspector-executor code,'' \emph{Proceedings of
  the IEEE}, vol. 106, no.~11, pp. 1921--1934, 2018.

\bibitem{blackford2002updated}
L.~S. Blackford, A.~Petitet, R.~Pozo, K.~Remington, R.~C. Whaley, J.~Demmel,
  J.~Dongarra, I.~Duff, S.~Hammarling, G.~Henry \emph{et~al.}, ``An updated set
  of basic linear algebra subprograms (blas),'' \emph{ACM Transactions on
  Mathematical Software}, vol.~28, no.~2, pp. 135--151, 2002.

\bibitem{davis2011university}
T.~A. Davis and Y.~Hu, ``The university of florida sparse matrix collection,''
  \emph{ACM Transactions on Mathematical Software (TOMS)}, vol.~38, no.~1, pp.
  1--25, 2011.

\bibitem{Davis2011}
\BIBentryALTinterwordspacing
------, ``{The University of Florida Sparse Matrix Collection},'' \emph{ACM
  Transactions on Mathematical Software}, vol.~38, no.~1, pp. 1--25, nov 2011.
  [Online]. Available: \url{https://dl.acm.org/doi/10.1145/2049662.2049663}
\BIBentrySTDinterwordspacing

\bibitem{Cheshmi2019}
K.~Cheshmi, S.~Kamil, M.~M. Strout, and M.~M. Dehnavi, ``{ParSy: Inspection and
  transformation of sparse matrix computations for parallelism},'' in
  \emph{Proceedings - International Conference for High Performance Computing,
  Networking, Storage, and Analysis, SC 2018}.\hskip 1em plus 0.5em minus
  0.4em\relax Institute of Electrical and Electronics Engineers Inc., mar 2019,
  pp. 779--793.

\bibitem{mkl_lib}
E.~Wang, Q.~Zhang, B.~Shen, G.~Zhang, X.~Lu, Q.~Wu, and Y.~Wang, \emph{Intel
  Math Kernel Library}, 05 2014, pp. 167--188.

\bibitem{davis2005cholmod}
T.~A. Davis and W.~Hager, ``Cholmod: supernodal sparse cholesky factorization
  and update/downdate,'' 2005.

\bibitem{hennessy_quantitative}
J.~Hennessy and D.~Patterson, \emph{Computer Architecture - A Quantitative
  Approach}, 01 2007.

\bibitem{terpstra2010collecting}
D.~Terpstra, H.~Jagode, H.~You, and J.~Dongarra, ``Collecting performance data
  with papi-c,'' in \emph{Tools for High Performance Computing 2009}.\hskip 1em
  plus 0.5em minus 0.4em\relax Springer, 2010, pp. 157--173.

\bibitem{chen2008algorithm}
Y.~Chen, T.~A. Davis, W.~W. Hager, and S.~Rajamanickam, ``Algorithm 887:
  Cholmod, supernodal sparse cholesky factorization and update/downdate,''
  \emph{ACM Transactions on Mathematical Software (TOMS)}, vol.~35, no.~3, pp.
  1--14, 2008.

\bibitem{benzi2000robust}
M.~Benzi, J.~K. Cullum, and M.~Tuma, ``Robust approximate inverse
  preconditioning for the conjugate gradient method,'' \emph{SIAM Journal on
  Scientific Computing}, vol.~22, no.~4, pp. 1318--1332, 2000.

\bibitem{kershaw1978incomplete}
D.~S. Kershaw, ``The incomplete cholesky-conjugate gradient method for the
  iterative solution of systems of linear equations,'' \emph{Journal of
  computational physics}, vol.~26, no.~1, pp. 43--65, 1978.

\bibitem{papadrakakis1993accuracy}
M.~Papadrakakis and N.~Bitoulas, ``Accuracy and effectiveness of preconditioned
  conjugate gradient algorithms for large and ill-conditioned problems,''
  \emph{Computer methods in applied mechanics and engineering}, vol. 109, no.
  3-4, pp. 219--232, 1993.

\bibitem{eigenweb}
G.~Guennebaud, B.~Jacob \emph{et~al.}, ``Eigen v3,''
  http://eigen.tuxfamily.org, 2010.

\bibitem{Li2003}
\BIBentryALTinterwordspacing
X.~S. Li and J.~W. Demmel, ``{SuperLU\_DIST: A scalable distributed-memory
  sparse direct solver for unsymmetric linear systems},'' \emph{ACM
  Transactions on Mathematical Software}, vol.~29, no.~2, pp. 110--140, jun
  2003. [Online]. Available: \url{https://dl.acm.org/doi/10.1145/779359.779361}
\BIBentrySTDinterwordspacing

\bibitem{Hutchinson1995}
S.~A. Hutchinson, J.~N. Shadid, and R.~S. Tuminaro, ``{Aztec User's Guide
  Version 1.1},'' Tech. Rep., 1995.

\bibitem{davis2004algorithm}
T.~A. Davis, ``Algorithm 832: Umfpack v4. 3---an unsymmetric-pattern
  multifrontal method,'' \emph{ACM Transactions on Mathematical Software
  (TOMS)}, vol.~30, no.~2, pp. 196--199, 2004.

\bibitem{li2013gpu}
R.~Li and Y.~Saad, ``Gpu-accelerated preconditioned iterative linear solvers,''
  \emph{The Journal of Supercomputing}, vol.~63, no.~2, pp. 443--466, 2013.

\bibitem{williams2007optimization}
S.~Williams, L.~Oliker, R.~Vuduc, J.~Shalf, K.~Yelick, and J.~Demmel,
  ``Optimization of sparse matrix-vector multiplication on emerging multicore
  platforms,'' in \emph{SC'07: Proceedings of the 2007 ACM/IEEE Conference on
  Supercomputing}.\hskip 1em plus 0.5em minus 0.4em\relax IEEE, 2007, pp.
  1--12.

\bibitem{kamin2014optimization}
S.~Kamin, M.~J. Garzar{\'a}n, B.~Aktemur, D.~Xu, B.~Y{\i}lmaz, and Z.~Chen,
  ``Optimization by runtime specialization for sparse matrix-vector
  multiplication,'' in \emph{Proceedings of the 2014 International Conference
  on Generative Programming: Concepts and Experiences}, 2014, pp. 93--102.

\bibitem{merrill2016merge}
D.~Merrill and M.~Garland, ``Merge-based parallel sparse matrix-vector
  multiplication,'' in \emph{SC'16: Proceedings of the International Conference
  for High Performance Computing, Networking, Storage and Analysis}.\hskip 1em
  plus 0.5em minus 0.4em\relax IEEE, 2016, pp. 678--689.

\bibitem{vuduc2003automatic}
R.~W. Vuduc, \emph{Automatic performance tuning of sparse matrix
  kernels}.\hskip 1em plus 0.5em minus 0.4em\relax University of California,
  Berkeley, 2003.

\bibitem{vuduc2005fast}
R.~W. Vuduc and H.-J. Moon, ``Fast sparse matrix-vector multiplication by
  exploiting variable block structure,'' in \emph{International Conference on
  High Performance Computing and Communications}.\hskip 1em plus 0.5em minus
  0.4em\relax Springer, 2005, pp. 807--816.

\bibitem{li2020vbsf}
Y.~Li, P.~Xie, X.~Chen, J.~Liu, B.~Yang, S.~Li, C.~Gong, X.~Gan, and H.~Xu,
  ``Vbsf: a new storage format for simd sparse matrix--vector multiplication on
  modern processors,'' \emph{The Journal of Supercomputing}, vol.~76, no.~3,
  pp. 2063--2081, 2020.

\bibitem{tang2015optimizing}
W.~T. Tang, R.~Zhao, M.~Lu, Y.~Liang, H.~P. Huyng, X.~Li, and R.~S.~M. Goh,
  ``Optimizing and auto-tuning scale-free sparse matrix-vector multiplication
  on intel xeon phi,'' in \emph{2015 IEEE/ACM International Symposium on Code
  Generation and Optimization (CGO)}.\hskip 1em plus 0.5em minus 0.4em\relax
  IEEE, 2015, pp. 136--145.

\bibitem{kreutzer2014unified}
M.~Kreutzer, G.~Hager, G.~Wellein, H.~Fehske, and A.~R. Bishop, ``A unified
  sparse matrix data format for efficient general sparse matrix-vector
  multiplication on modern processors with wide simd units,'' \emph{SIAM
  Journal on Scientific Computing}, vol.~36, no.~5, pp. C401--C423, 2014.

\bibitem{davis2010algorithm}
T.~A. Davis and E.~Palamadai~Natarajan, ``Algorithm 907: Klu, a direct sparse
  solver for circuit simulation problems,'' \emph{ACM Transactions on
  Mathematical Software (TOMS)}, vol.~37, no.~3, pp. 1--17, 2010.

\bibitem{jones1993partial}
N.~D. Jones, C.~K. Gomard, and P.~Sestoft, \emph{Partial evaluation and
  automatic program generation}.\hskip 1em plus 0.5em minus 0.4em\relax Peter
  Sestoft, 1993.

\bibitem{quillere2000generation}
F.~Quiller{\'e}, S.~Rajopadhye, and D.~Wilde, ``Generation of efficient nested
  loops from polyhedra,'' \emph{International journal of parallel programming},
  vol.~28, no.~5, pp. 469--498, 2000.

\bibitem{spampinato2014basic}
D.~G. Spampinato and M.~P{\"u}schel, ``A basic linear algebra compiler,'' in
  \emph{Proceedings of Annual IEEE/ACM International Symposium on Code
  Generation and Optimization}, 2014, pp. 23--32.

\bibitem{alnaes2014unified}
M.~S. Aln{\ae}s, A.~Logg, K.~B. {\O}lgaard, M.~E. Rognes, and G.~N. Wells,
  ``Unified form language: A domain-specific language for weak formulations of
  partial differential equations,'' \emph{ACM Transactions on Mathematical
  Software (TOMS)}, vol.~40, no.~2, pp. 1--37, 2014.

\bibitem{luporini2017algorithm}
F.~Luporini, D.~A. Ham, and P.~H. Kelly, ``An algorithm for the optimization of
  finite element integration loops,'' \emph{ACM Transactions on Mathematical
  Software (TOMS)}, vol.~44, no.~1, pp. 1--26, 2017.

\bibitem{Shin2010}
\BIBentryALTinterwordspacing
J.~Shin, M.~W. Hall, J.~Chame, C.~Chen, P.~F. Fischer, and P.~D. Hovland,
  ``{Speeding up Nek5000 with autotuning and specialization},'' in
  \emph{Proceedings of the International Conference on Supercomputing}.\hskip
  1em plus 0.5em minus 0.4em\relax New York, New York, USA: ACM Press, 2010,
  pp. 253--262. [Online]. Available:
  \url{http://portal.acm.org/citation.cfm?doid=1810085.1810120}
\BIBentrySTDinterwordspacing

\bibitem{Tang2011}
\BIBentryALTinterwordspacing
Y.~Tang, R.~A. Chowdhury, B.~C. Kuszmaul, C.~K. Luk, and C.~E. Leiserson,
  ``{The pochoir stencil compiler},'' in \emph{Annual ACM Symposium on
  Parallelism in Algorithms and Architectures}.\hskip 1em plus 0.5em minus
  0.4em\relax New York, New York, USA: ACM Press, 2011, pp. 117--128. [Online].
  Available: \url{http://portal.acm.org/citation.cfm?doid=1989493.1989508}
\BIBentrySTDinterwordspacing

\bibitem{VanDerSpek2011}
H.~L.~A. van~der Spek and H.~A.~G. Wijshoff, ``Sublimation: Expanding data
  structures to enable data instance specific optimizations,'' in
  \emph{Languages and Compilers for Parallel Computing}, K.~Cooper,
  J.~Mellor-Crummey, and V.~Sarkar, Eds.\hskip 1em plus 0.5em minus 0.4em\relax
  Berlin, Heidelberg: Springer Berlin Heidelberg, 2011, pp. 106--120.

\bibitem{vasilache2006polyhedral}
N.~Vasilache, C.~Bastoul, and A.~Cohen, ``Polyhedral code generation in the
  real world,'' in \emph{International Conference on Compiler
  Construction}.\hskip 1em plus 0.5em minus 0.4em\relax Springer, 2006, pp.
  185--201.

\bibitem{agrawal1995interprocedural}
G.~Agrawal, J.~Saltz, and R.~Das, ``Interprocedural partial redundancy
  elimination and its application to distributed memory compilation,''
  \emph{ACM SIGPLAN Notices}, vol.~30, no.~6, pp. 258--269, 1995.

\bibitem{das1995index}
R.~Das, P.~Havlak, J.~Saltz, and K.~Kennedy, ``Index array flattening through
  program transformation,'' in \emph{Proceedings of the 1995 ACM/IEEE
  conference on Supercomputing}, 1995, pp. 70--es.

\bibitem{Rodriguez}
\BIBentryALTinterwordspacing
G.~Rodr{\'{i}}guez and L.~Pouchet, ``{Polyhedral Modeling of Immutable Sparse
  Matrices},'' \emph{impact.gforge.inria.fr}. [Online]. Available:
  \url{http://impact.gforge.inria.fr/impact2018/papers/modeling-immutable-sparsemat.pdf}
\BIBentrySTDinterwordspacing

\bibitem{saltz1990run}
J.~Saltz, K.~Crowley, R.~Michandaney, and H.~Berryman, ``Run-time scheduling
  and execution of loops on message passing machines,'' \emph{Journal of
  Parallel and Distributed Computing}, vol.~8, no.~4, pp. 303--312, 1990.

\bibitem{lamielle2010enabling}
A.~LaMielle and M.~M. Strout, ``Enabling code generation within the sparse
  polyhedral framework,'' \emph{Technical report, Technical Report CS-10-102},
  2010.

\bibitem{venkat2016automating}
A.~Venkat, M.~S. Mohammadi, J.~Park, H.~Rong, R.~Barik, M.~M. Strout, and
  M.~Hall, ``Automating wavefront parallelization for sparse matrix
  computations,'' in \emph{SC'16: Proceedings of the International Conference
  for High Performance Computing, Networking, Storage and Analysis}.\hskip 1em
  plus 0.5em minus 0.4em\relax IEEE, 2016, pp. 480--491.

\bibitem{Rodriguez2016}
\BIBentryALTinterwordspacing
G.~Rodr{\'{i}}guez, J.~M. Andi{\'{o}}n, M.~T. Kandemir, and J.~Touri{\~{n}}o,
  ``{Trace-based affine reconstruction of codes},'' in \emph{Proceedings of the
  14th International Symposium on Code Generation and Optimization, CGO
  2016}.\hskip 1em plus 0.5em minus 0.4em\relax New York, NY, USA: Association
  for Computing Machinery, Inc, feb 2016, pp. 139--149. [Online]. Available:
  \url{https://dl.acm.org/doi/10.1145/2854038.2854056}
\BIBentrySTDinterwordspacing

\bibitem{Ketterlin2008}
\BIBentryALTinterwordspacing
A.~Ketterlin and P.~Clauss, ``{Prediction and trace compression of data access
  addresses through nested loop recognition},'' in \emph{Proceedings of the
  2008 CGO - Sixth International Symposium on Code Generation and
  Optimization}.\hskip 1em plus 0.5em minus 0.4em\relax New York, New York,
  USA: ACM Press, 2008, pp. 94--103. [Online]. Available:
  \url{http://portal.acm.org/citation.cfm?doid=1356058.1356071}
\BIBentrySTDinterwordspacing

\bibitem{Selva2019}
\BIBentryALTinterwordspacing
M.~Selva, F.~Gruber, D.~Sampaio, C.~Guillon, L.~N. Pouchet, and F.~Rastello,
  ``{Building a polyhedral representation from an instrumented execution:
  Making dynamic analyses of nonaffine programs scalable},'' \emph{ACM
  Transactions on Architecture and Code Optimization}, vol.~16, no.~4, pp.
  1--26, dec 2019. [Online]. Available:
  \url{https://dl.acm.org/doi/10.1145/3363785}
\BIBentrySTDinterwordspacing

\end{thebibliography}

\end{document}